\documentclass[preprint,12pt]{elsarticle}

\usepackage{amssymb}

\usepackage{lipsum}
\usepackage[utf8]{inputenc}
\usepackage{epstopdf}
\usepackage{algorithmic}
\ifpdf
  \DeclareGraphicsExtensions{.eps,.pdf,.png,.jpg}
\else
  \DeclareGraphicsExtensions{.eps}
\fi
\usepackage[caption=false]{subfig}
\usepackage{amsmath,amssymb,amsfonts}
\usepackage{verbatim}
\usepackage{xcolor}
\DeclareMathOperator{\sech}{sech}
\newcommand\T{\rule{0pt}{2.6ex}}       
\newcommand\B{\rule[-1.2ex]{0pt}{0pt}} 
\newcommand{\rmd}{\mathrm{d}}

\begin{document}

\begin{frontmatter}



\title{Analysis of BBM solitary wave interactions using the conserved quantities\tnoteref{label1}}
\tnotetext[label1]{Published in \emph{Chaos, Solitons and Fractals} (10.1016/j.chaos.2021.111725).}


\author[inst1]{Xiangcheng You}

\affiliation[inst1]{organization={School of Petroleum Engineering, China University of Petroleum-Beijing},
            city={Beijing},
            postcode={102249}, 
            country={China}}

\author[inst2]{Hang Xu}
\affiliation[inst2]{organization={State Key Laboratory of Ocean Engineering, School of Naval Architecture Ocean and Civil Engineering, Shanghai Jiao Tong University},
            city={Shanghai},
            postcode={200240}, 
            country={China}}

\author[inst3]{Qiang Sun\corref{cor1}}
\cortext[cor1]{Corresponding author: qiang.sun@rmit.edu.au}
\affiliation[inst3]{organization={Australian Research Council Centre of Excellence for Nanoscale BioPhotonics, School of Science, RMIT University},
            city={Melbourne},
            postcode={VIC 3001}, 
            country={Australia}}

\begin{abstract}
In this paper, a simple, robust, fast and effective method based on the conserved quantities is developed to approximate and analyse the shape, structure and interaction characters of the solitary waves described by the Benjamin-Bona-Mahony (BBM) equation. Due to the invariant character of the conserved quantities, there is no need to solve the related complex nonlinear partial differential BBM equation to simulate the interactions between the solitary waves at the most merging instance. Good accuracy of the proposed method has been found when compared with the numerical method for the solitary wave interactions with different initial incoming wave shapes. The conserved quantity method developed in
this work can serve as an ideal tool to benchmark numerical solvers, to perform the stability analysis, and to analyse the interacting phenomena between solitary waves.
\end{abstract}


\begin{highlights}
\item Developed a simple approach based on the conserved quantities to analyse the solitary wave interactions described by the Benjamin-Bona-Mahony (BBM) equation.
\item The main advantage of this approach is that it does not need to solve the nonlinear partial differential BBM equation when simulating the interactions between the solitary waves at the most merging instance.
\item Good agreement is found between the results of the current method and the numerical results.
\item This method is ideal to benchmark numerical solvers, perform stability analysis, and analyse soliton interactions for shallow water waves.
\end{highlights}

\begin{keyword}
KdV solitary waves \sep interactions \sep integral invariants
\end{keyword}

\end{frontmatter}


\section{Introduction}

The interaction between surface solitary waves in a nonlinear dispersive system can be seen in
many different areas, such as fluid mechanics, plasma physics, neuro physics, solid-state physics and
nonlinear optics \cite{Salupere2003137,Kochanov201325,Roshid201749,Khater20211797,Han2019236}.
The Korteweg de Vries (KdV) equation and its substitute model, the Benjamin-Bona-Mahony (BBM) equation \cite{Benjamin197247},
can be used to study the long surface solitary waves with finite amplitude
propagating unidirectionally in a nonlinear dispersive medium. Relative to the KdV model, the BBM
equation is reconganized as a regularized version of KdV equation with two properties: there are only
three integrals of the wave motion and its solution is stable at high wavenumbers.

The BBM equation is a nonlinear partial differential equation. To find analytical solutions of the BBM equation
is of great importance, since analytical solutions can be used to benchmark numerical
solvers, to perform the stability analysis, and to grasp a better understanding of the mechanism of
complicated physics phenomena. Various direct methods have been developed along this line, such
as the tanh method \cite{Wazwaz20081505}, the Lie group analysis \cite{Molati2012212},
the (G'/G)-expansion method \cite{Manafianheris201228}, the exp-function
and F-expansion method \cite{Triki201048}, the auxiliary equation method \cite{Khater20191}, the homogeneous balance method \cite{Gomez20101430}
and so on \cite{Khalique2013413,Ren2021104159}. Also, numerical methods have been developed to solve the BBM
equation, for example the finite element method \cite{Karakoc20191917,Wang2021168}, the finite difference method \cite{Omrani2008999,Shi201674}, the spectral method \cite{Sloan1991159}
and the B-spline collocation method \cite{Karakoc2014596406}.

In general, the solution procedures of those direct solution methods for solving the BBM equation are complicated, and the forms of the obtained solutions are usually too complex to analyse, in
particular for the problems of the interaction between solitary waves. Nevertheless, the solitary wave
interactions are of interests not only in fundamental research but also in practical applications
\cite{Zabusky1965240,Su1980509,Craig200657106,Chambarel2009111}.
As such, it is desired to have a simple, robust, fast and effective analysis method
to study the interactions of multiple solitary waves possessed by the BBM equation.

To find such a method, one of the property of the BBM equation gets our attention. There are
only three integrals of the wave motion possessed by the BBM equation, and these three integrals, the
integral invariants or conserved quantities, represent the mass, the momentum and the energy of the wave. It
is well known that the collision of two solitons is elastic in integrable models, which means that they interact
without emitting any radiation, while in non-integrable models the interactions are usually nearly elastic~\cite{Bogolubsky1977149},
for example the solitary waves of the BBM equation. As such, when the wave travels in a stationary
reference frame, the three conserved quantities remain constant over time that represents the mass,
the momentum, and the energy. If two incoming solitary waves interact, these quantities can be added up
correspondingly to get the total mass, momentum and energy of the merged wave. Based on
this idea, we propose a simple, robust, fast and effective method by using the conserved quantities
to study the interaction between two solitary waves of the BBM equation at their maximum merging
moment. Such an idea has been used to simulate the nonlinear evolution of Gaussian waves group in
deep water \cite{Adcock20093083,Adcock2010569}. Also, a new local energy-preserving algorithm with the similar idea for the BBM
equation has also been proposed based on the temporal and spatial discretisations \cite{Yang2018119} in which the local mass
and the local energy are conserved in the local time-space region. Moreover, when using numerical methods to solve the nonlinear Schr\"{o}dinger equation with wave operator, it has been demonstrated that the numerical methods with conservative scheme can perform better than that without the conservative scheme~\cite{Li2018a,Li2018b,Li2019,Li2021} as more details in physical processes can be captured in the numerical solutions when the invariable properties of mass and energy are kept.

The structure of this work proceeds as follows. In Sec.2, the mathematical model is described and the three conserved quantities of the wave motion are defined. In Sec. 3,
the detailed conserved quantity method is demonstrated, and the analysis of three types of merging phenomena of two solitary waves of the BBM equation are presented. The conclusion is given in Sec. 4.

\section{Mathematical model}

As an improvement or a substitute model of the Korteweg de Vries (KdV) equation, the Benjamin-Bona-Mahony (BBM) equation (\ref{eq0001}) describes the long surface waves travelling unidirectionally in a nonlinear dispersive system. The BBM equation is also called as the regularized long-wave equation which is the form of
\begin{equation}
u_t+u_x+u u_x-u_{xxt}=0.\label{eq0001}
\end{equation}
In Eq. (\ref{eq0001}), $u$ is the non-dimensional surface elevation, $t$ the non-dimensional time and $x$ the non-dimensional displacement.
Also, $u_t=\partial u/\partial t$ and $u_x=\partial u/\partial x$.

The solitary wave solutions possessed by the BBM equation can be found by applying a general solitary wave shape as
\begin{equation}
u=\bar{a} \sech^2\left[\bar{b}(x-\bar{c}t)\right],\label{eq0002}
\end{equation}
where $\bar{a},\bar{b},\bar{c}$ are the coefficients related to the speed of wave.
When we let $\bar{a}=12b^2, \bar{b}=b/\sqrt{1+4b^2}, \bar{c}=1+4b^2$, and introduce Eq. (\ref{eq0002}) into Eq. (\ref{eq0001}), we can find that it satisfies the BBM
equation directly. As such, the solitary wave solutions to the BBM equation can be fully determined by one parameter which is the inverse width $b$, and the solution is
\begin{equation}
u=12b^2 \sech^2\left\{\frac{b[x-(1+4b^2)t]}{\sqrt{1+4b^2}}  \right\}.\label{eq0003}
\end{equation}

It has been proved that the BBM equation is not completely reversible because it has only three invariants which corresponds to the conservation of mass, momentum and energy. The
expressions of these three conserved quantities are, respectively,
\begin{align}
&I_1\equiv \int^{\infty}_{-\infty} u \, \rmd x,\label{eq0004_1}\\&
I_2\equiv \int^{\infty}_{-\infty} u^2 \, \rmd x,\label{eq0004_2}\\&
I_3\equiv \int^{\infty}_{-\infty} (u^3-3u_x^2) \, \rmd x.\label{eq0004_3}
\end{align}
It can be straightforwardly interpreted that the above three quantities represents mass ($I_1$), momentum
($I_2$) and energy ($I_3$), and they all remain constant along with time as the waves travel. Such a conserved property of $I_1$, $I_2$ and $I_3$ can be proved by integrating the BBM equation over the entire space of interest~\cite{Taylor2016}. For example,
\begin{align}\label{eq:I1proof}
    \int_{-\infty}^{\infty} u_t \,\rmd x = - \int_{-\infty}^{\infty} \frac{\partial }{\partial x}\left(u + \frac{1}{2}u^2 - u_{xt}\right) \,\rmd x = \left(u + \frac{1}{2}u^2 - u_{xt}\right) \bigg{|}_{-\infty}^{\infty}. 
\end{align}
Naturally, the wave motion decays to zero at a certain distance, so that the right-hand side of Eq.~(\ref{eq:I1proof}) equals to zero. Moving the time derivative out of the integration over space on the left-hand side of Eq.~(\ref{eq:I1proof}), we have
\begin{align}
    \frac{\partial }{\partial t}\int_{-\infty}^{\infty} u \,\rmd x = 0
\end{align}      
which leads to   
\begin{align}
    I_1\equiv \int^{\infty}_{-\infty} u \, \rmd x = \text{Constant}.
\end{align} 
In the similar manner, the higher order quantities, $I_2$ and $I_3$, can be derived from the partial differential equations of the BBM equation, and the conservation of $I_2$ and $I_3$ can be found in the meantime.

The quantities in Eqs.~(\ref{eq0004_1}) to~(\ref{eq0004_3}),
also known as the integral invariants, can be added with the invariants of another wave if the two merge,
to get the total mass, momentum and energy of the merged wave shape. Such a property provides
a simple, robust and efficient tool to study the interactions between solitary waves described by the
BBM equation. For example, when two waves with different amplitudes (and therefore different inverse
widths $b_1, b_2$) merge, these quantities conserve when they interact. As such, the total mass involved
in these two waves always remain the same. This means that at the moment when the waves merge
most, there will be a symmetric shape, and the sum of the two values of $I_1$ from the input waves with
parameters of $b_1$ and $b_2$ equals to that of the symmetric shape of the merged wave. Likewise for $I_2$ and
$I_3$. Taking advantage of these properties, various useful results can be obtained with the minimal
requirement of computational efforts, as shown in Sec. 3.

\section{BBM solitary waves interactions}

\begin{figure*}[!t]
    \centering{}
    \subfloat[]{\includegraphics[width=0.36\textwidth]{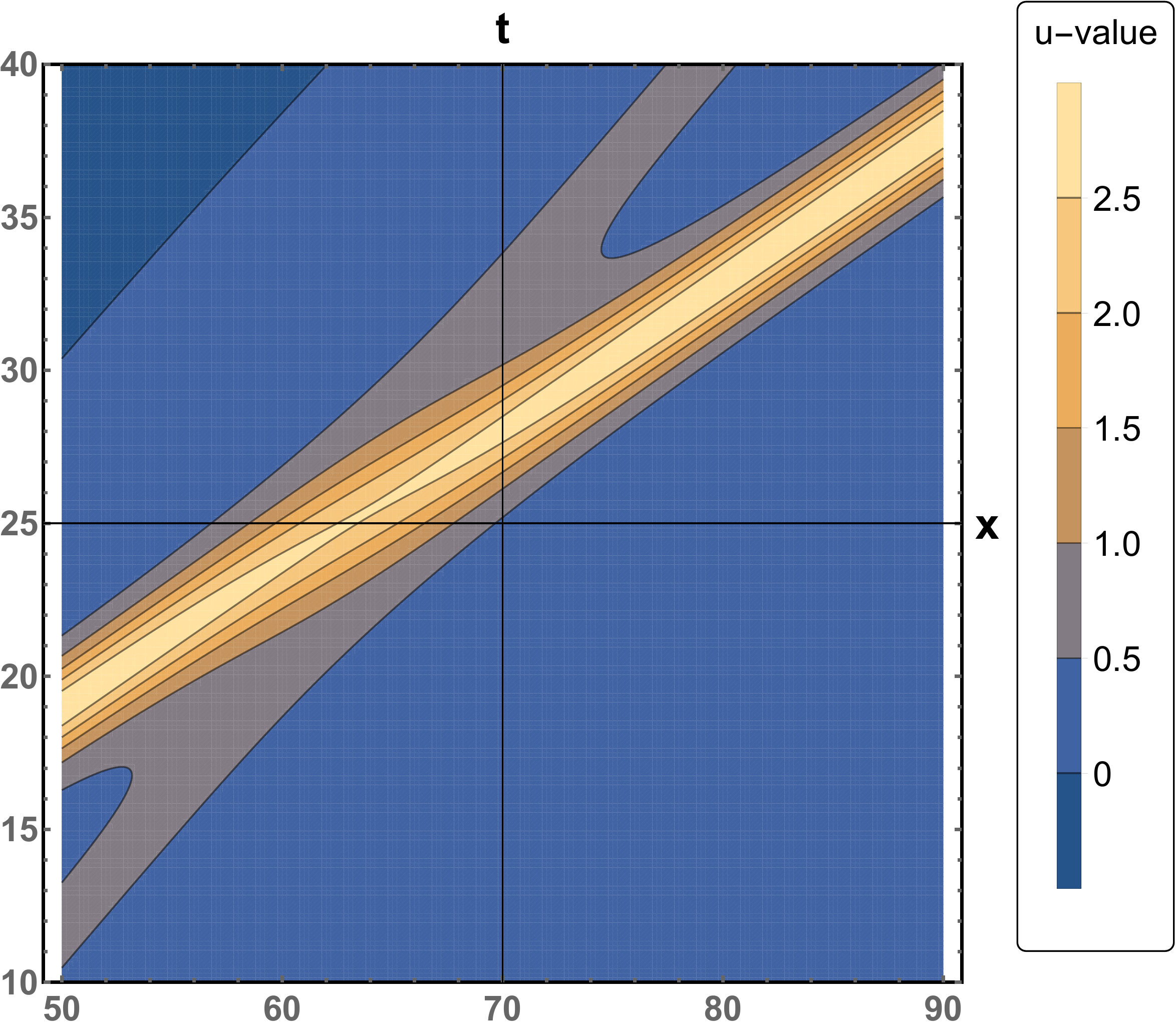}} \quad\quad
    \subfloat[]{\includegraphics[width=0.36\textwidth]{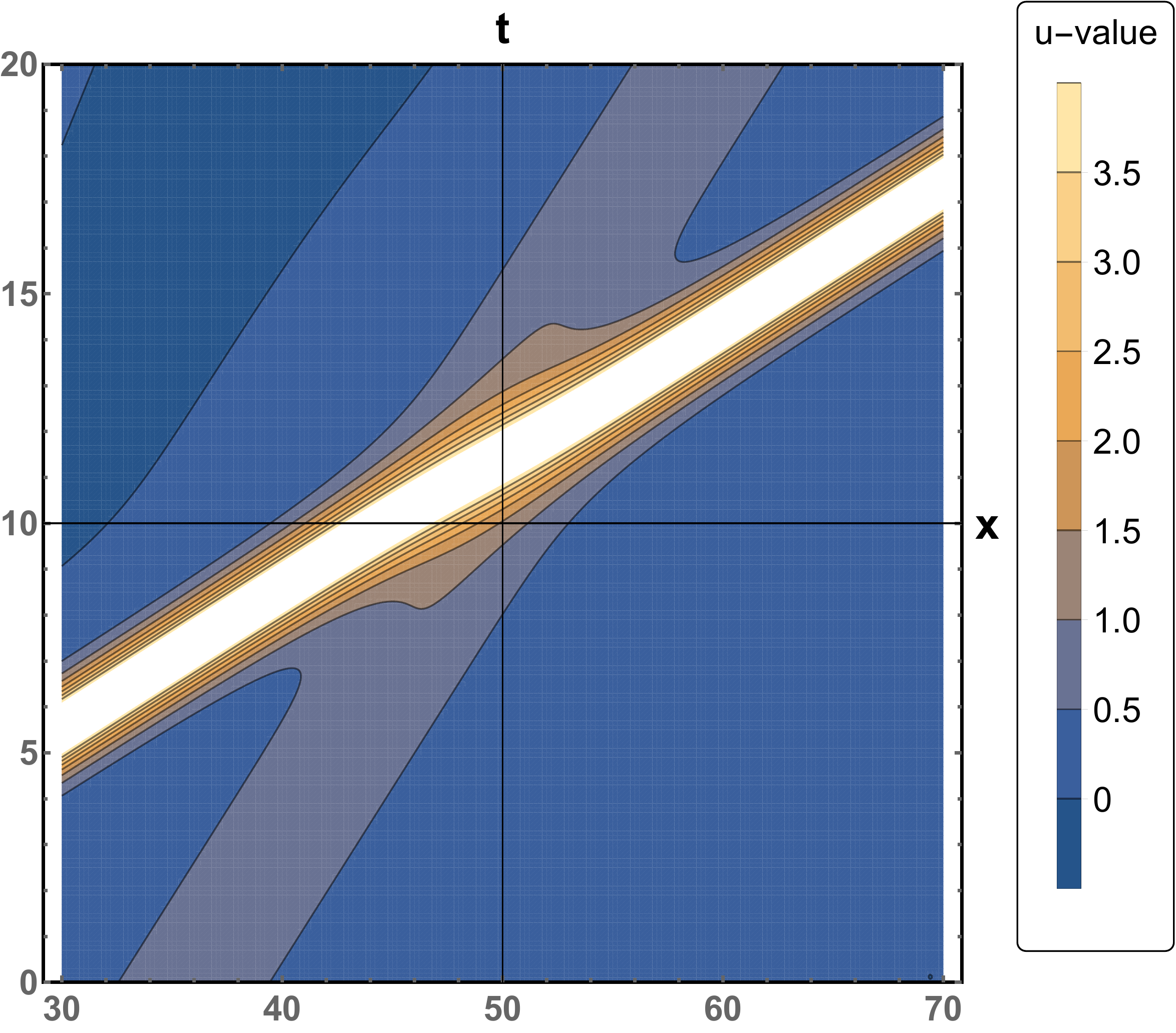}} \\ 
    \subfloat[]{\includegraphics[width=0.36\textwidth]{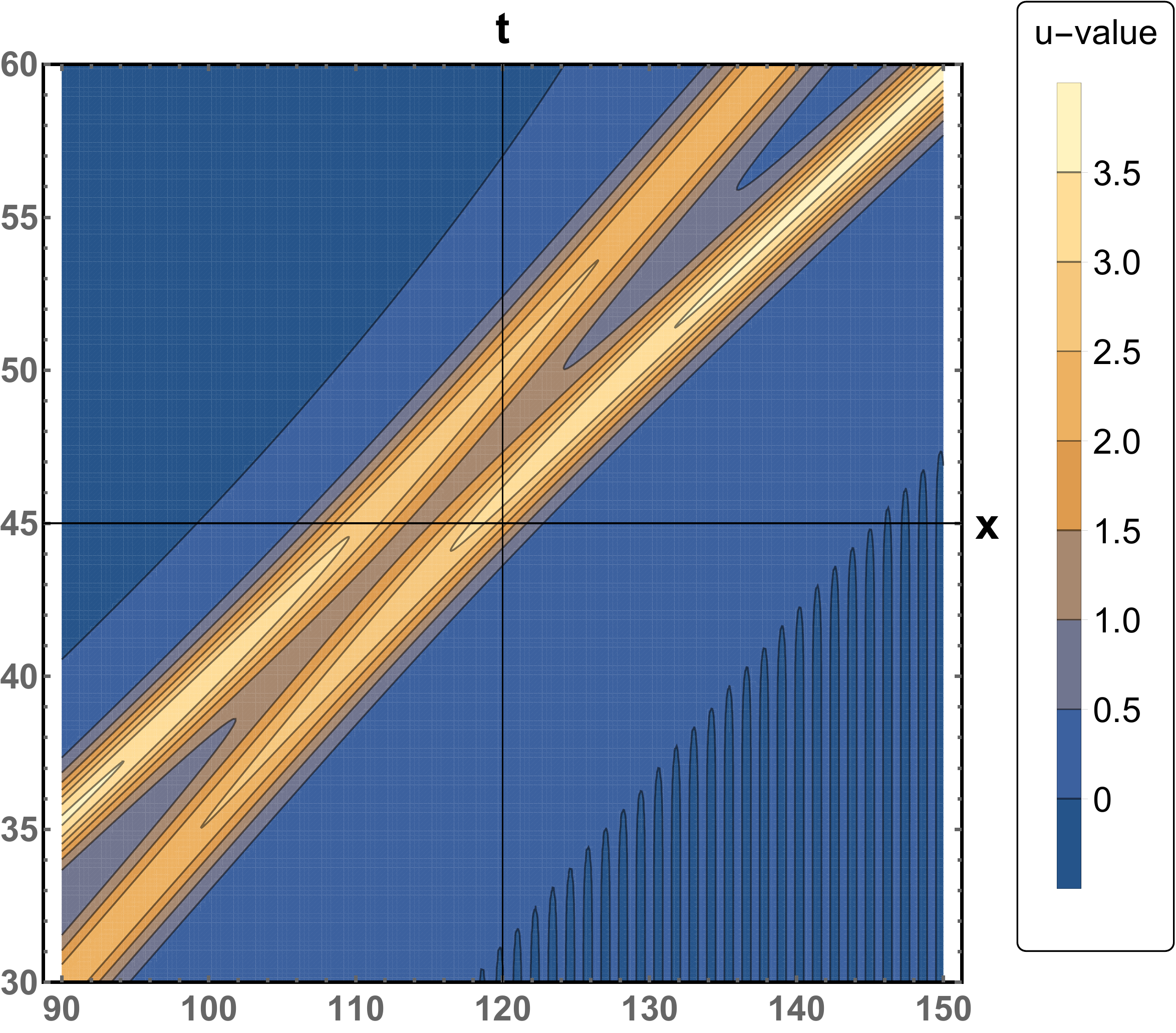}} \quad\quad
    \subfloat[]{\includegraphics[width=0.36\textwidth]{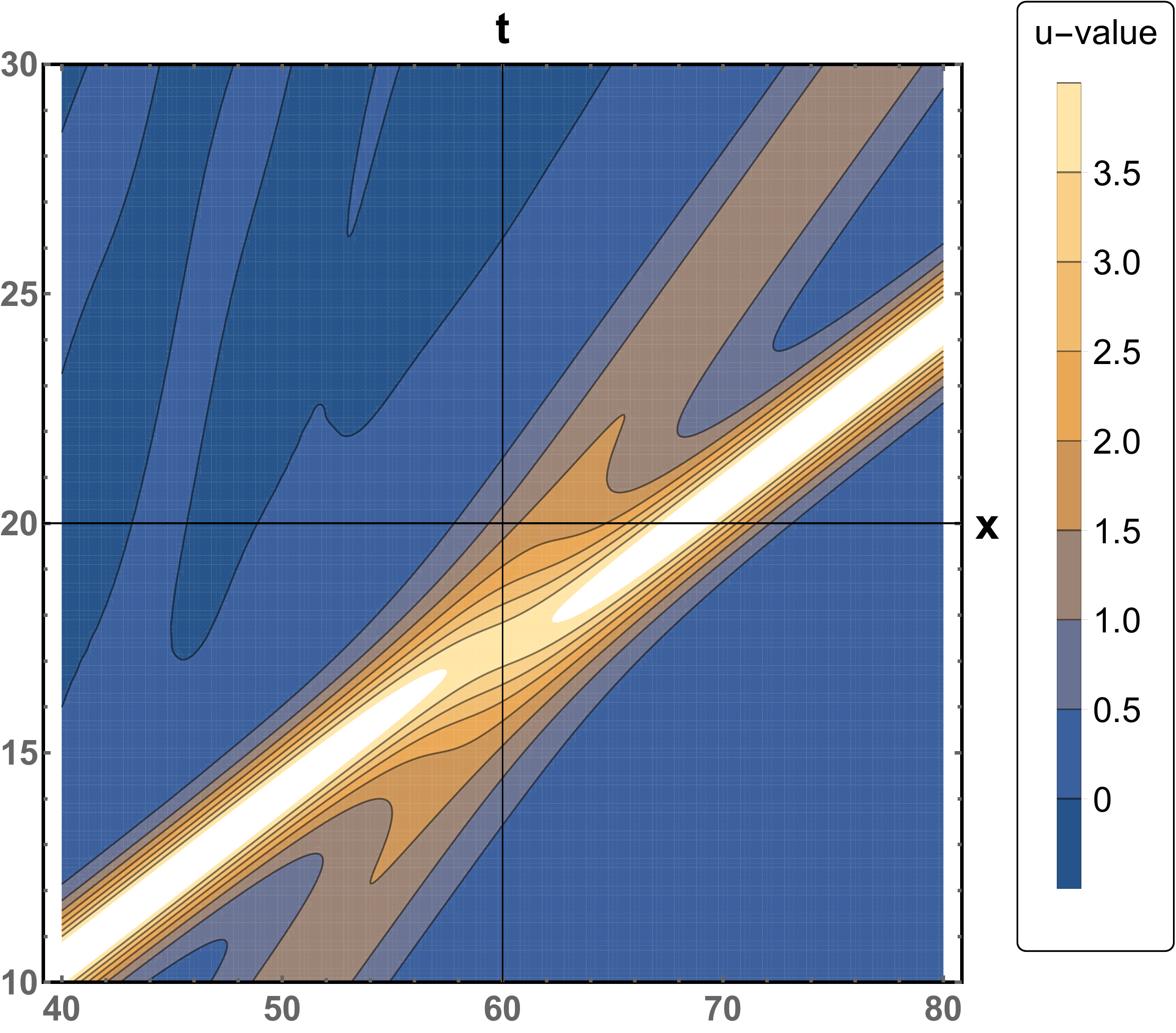}} \\
    \subfloat[]{\includegraphics[width=0.36\textwidth]{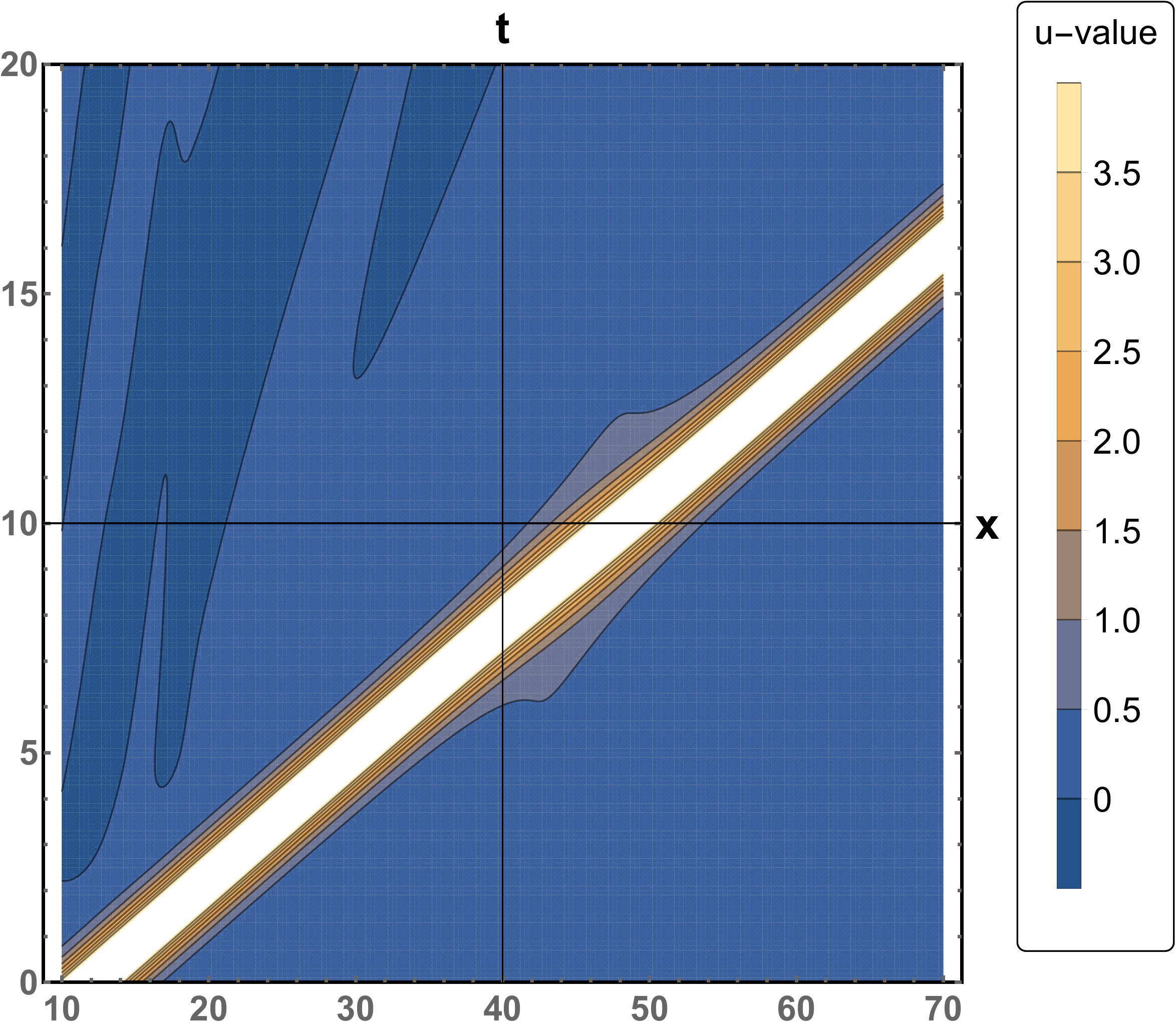}} \quad\quad
    \subfloat[]{\includegraphics[width=0.36\textwidth]{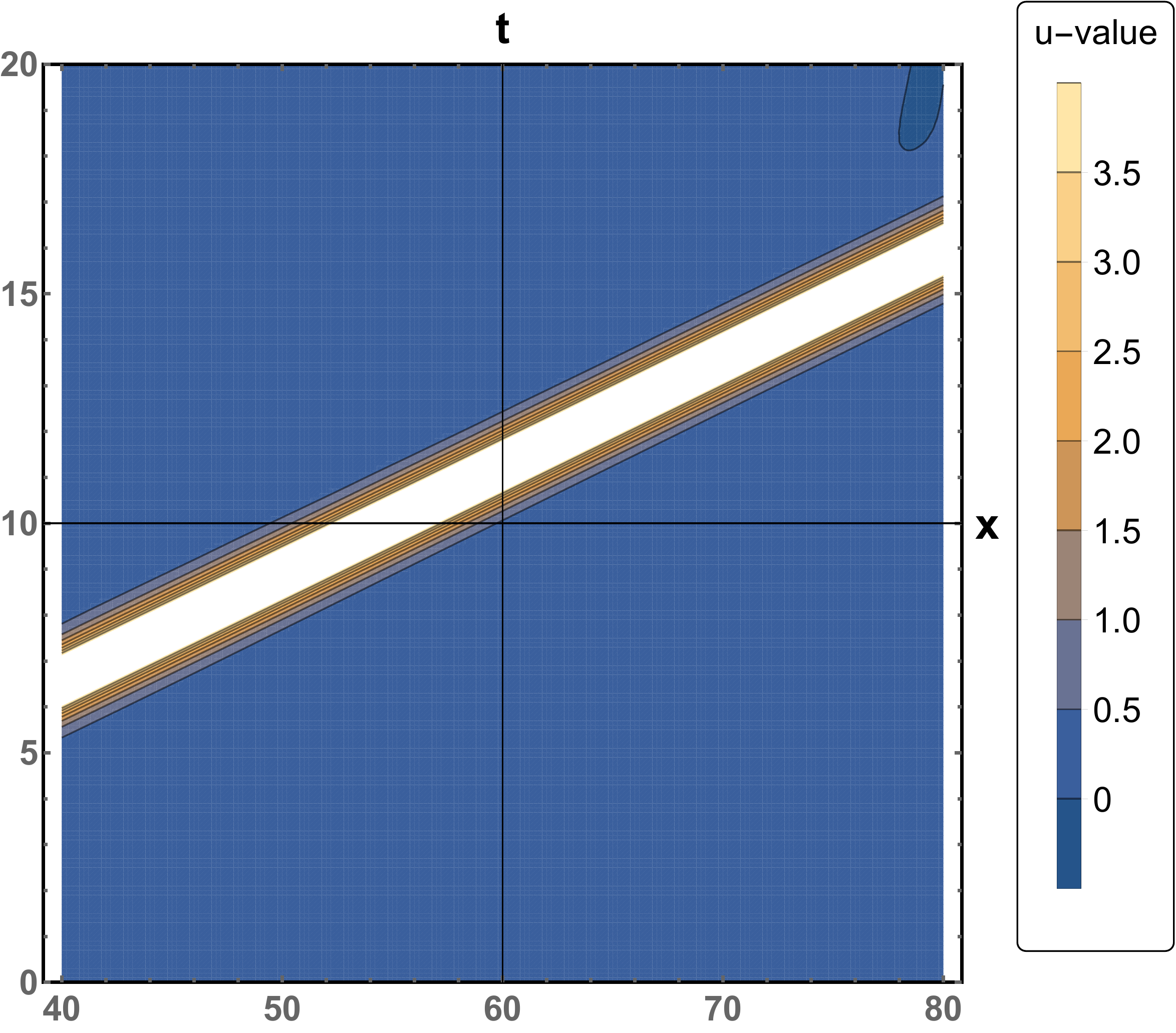}}
    \caption{Contour plots of interactions between solitons possessed by the BBM equation: (i) perfect merging interaction cases of with (a) $b_1=0.5, b_2=0.214299$ and (b) $b_1=0.75, b_2=0.280551$; (ii) peak merging interaction cases with (c) $b_1=0.55, b_2=0.45$ and (d) $b_1=0.65, b_2=0.35$; (iii) runover merging interaction cases with (e) $b_1=0.8, b_2=0.2$ and (f) $b_1=0.9, b_2=0.1$.}\label{fig1}
\end{figure*}

In this section, we demonstrate a few examples of the interactions between two solitary waves of
the BBM equation by using the integral invariants. This covers most practical interactions because
when more than two waves propagate, they usually travel at different speed and are likely to interact
as pairs at different time frames.

\begin{figure*}[!t]
    \centering{}
    \subfloat[]{\includegraphics[width=0.45\textwidth]{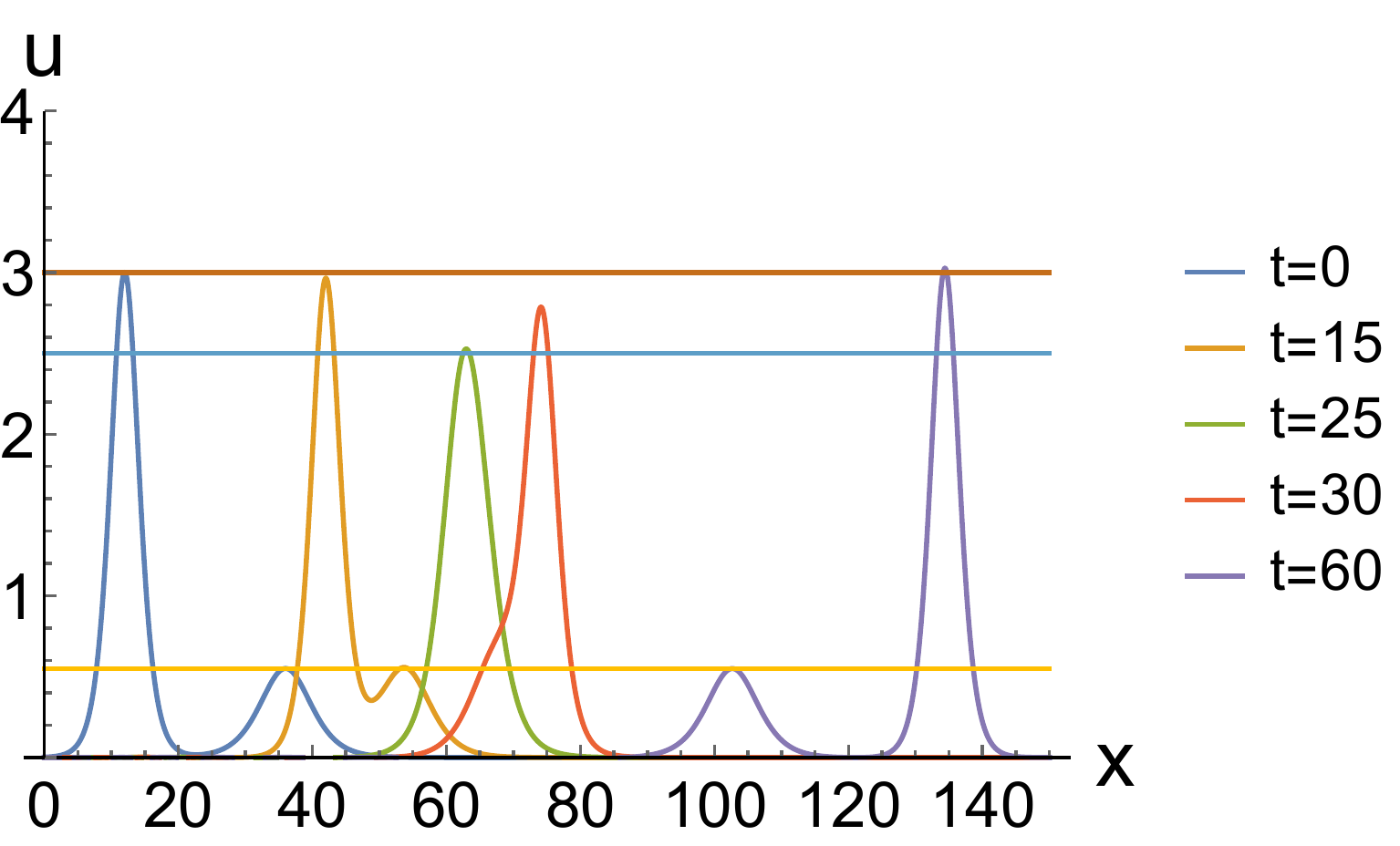}} \quad\quad
    \subfloat[]{\includegraphics[width=0.45\textwidth]{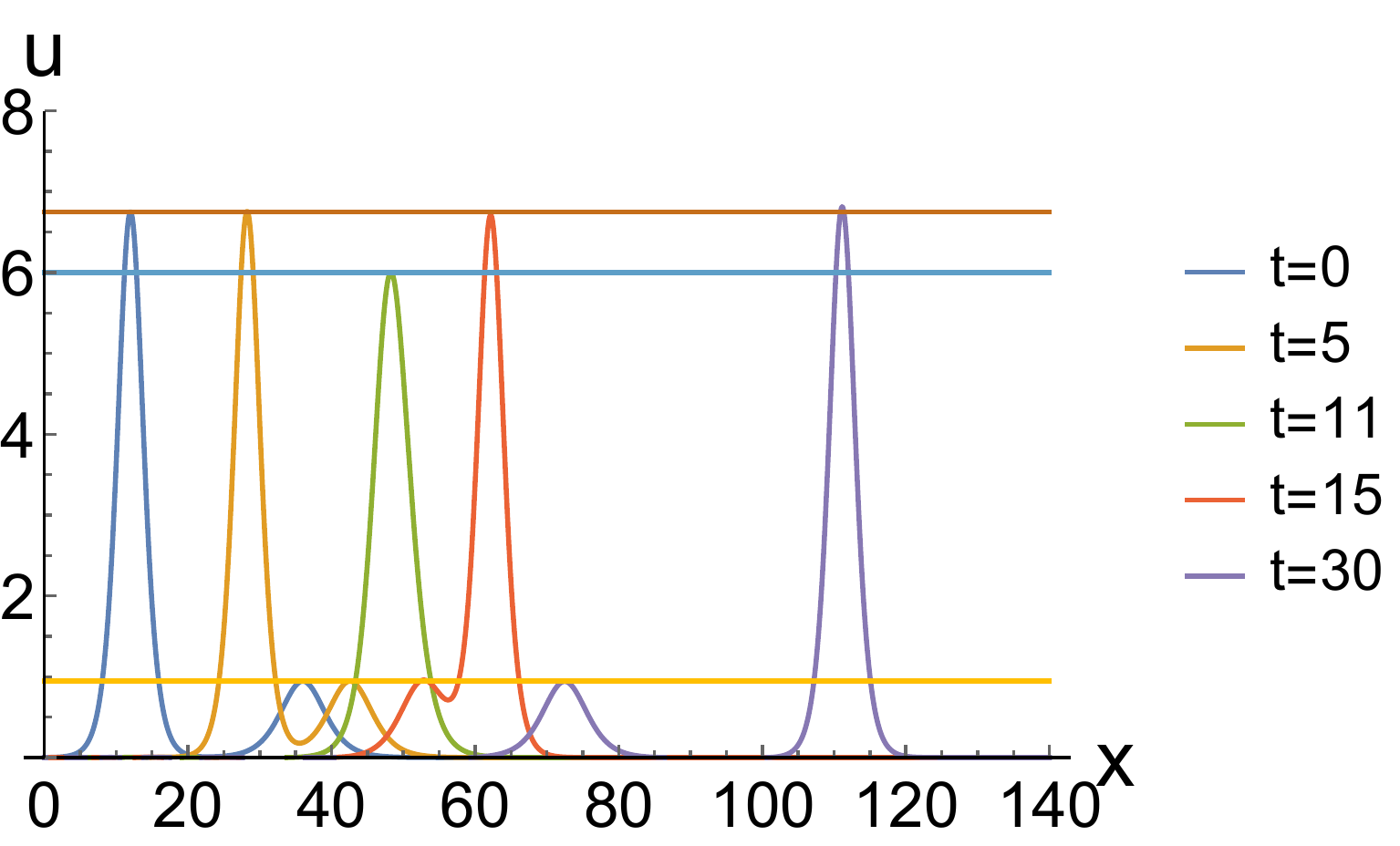}} \\ 
    \subfloat[]{\includegraphics[width=0.45\textwidth]{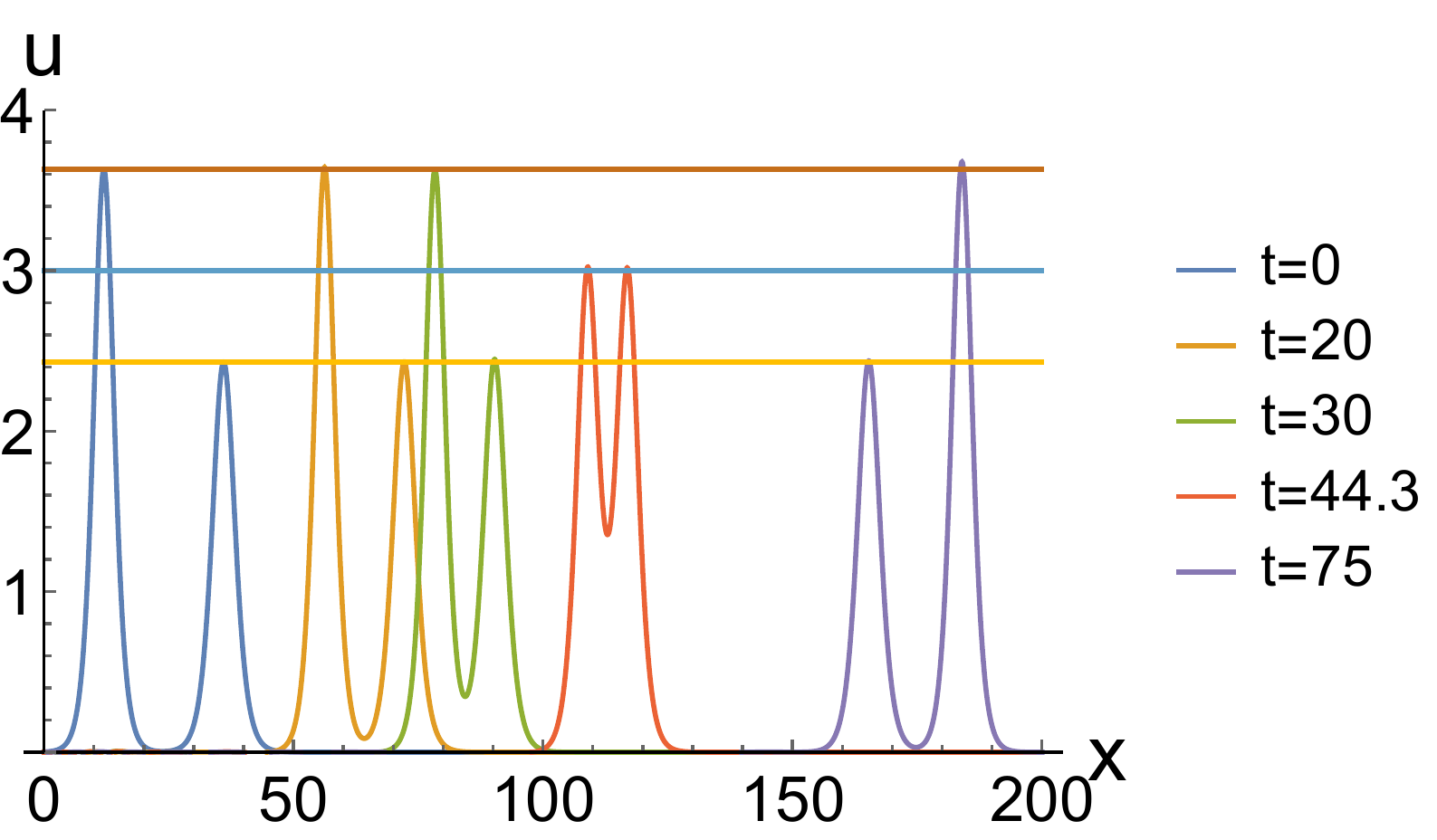}} \quad\quad
    \subfloat[]{\includegraphics[width=0.45\textwidth]{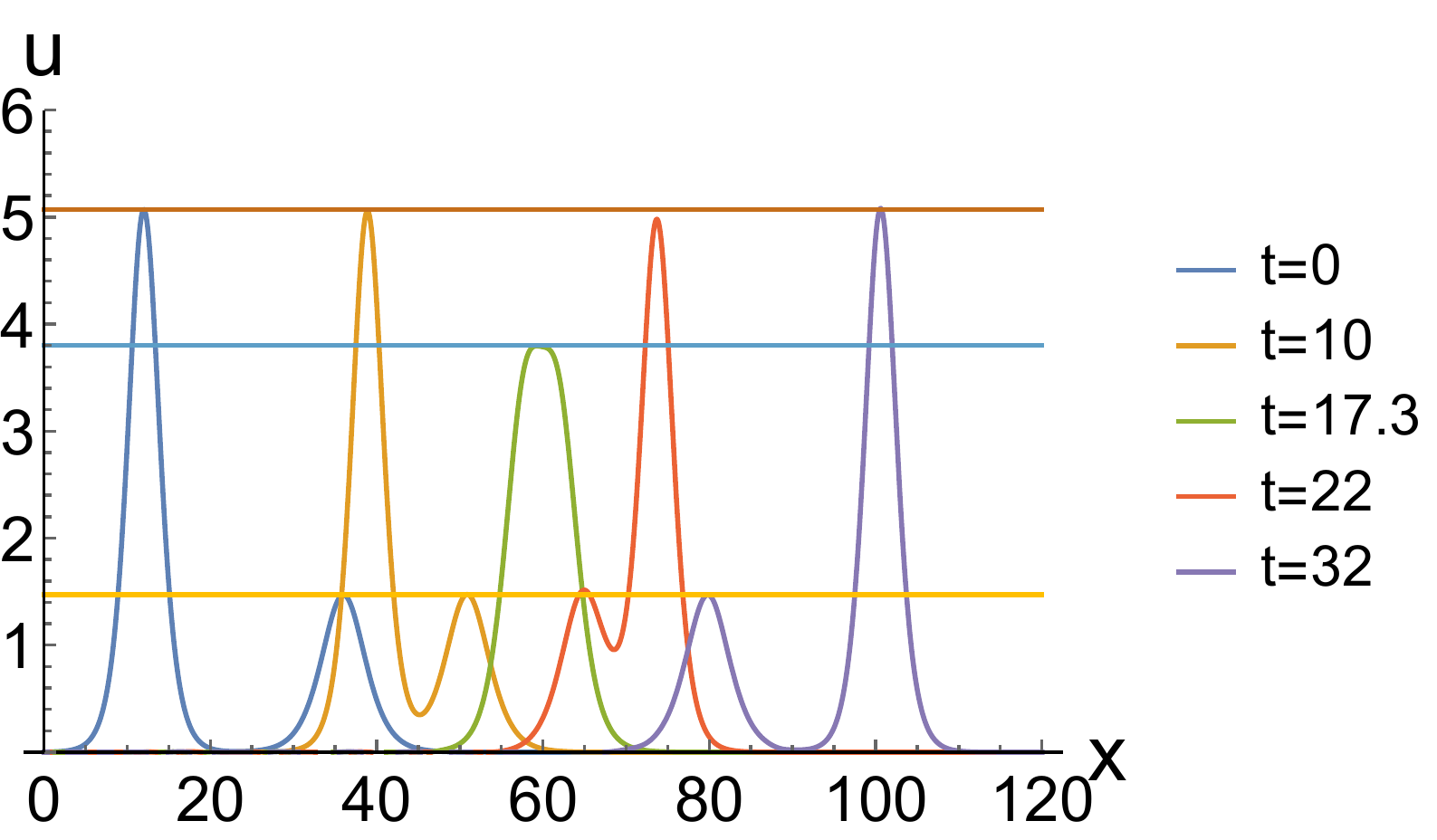}} \\
    \subfloat[]{\includegraphics[width=0.45\textwidth]{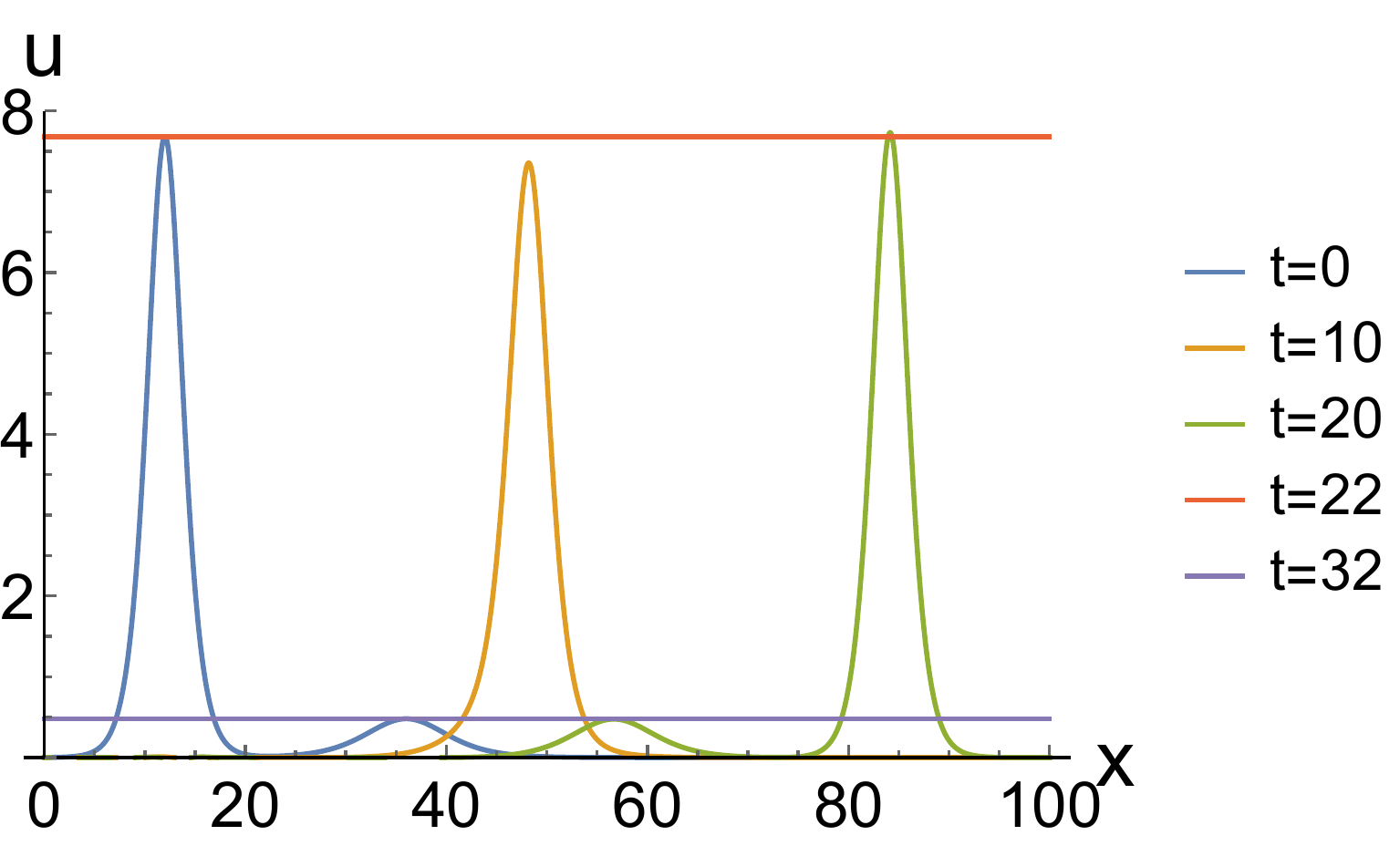}} \quad\quad
    \subfloat[]{\includegraphics[width=0.45\textwidth]{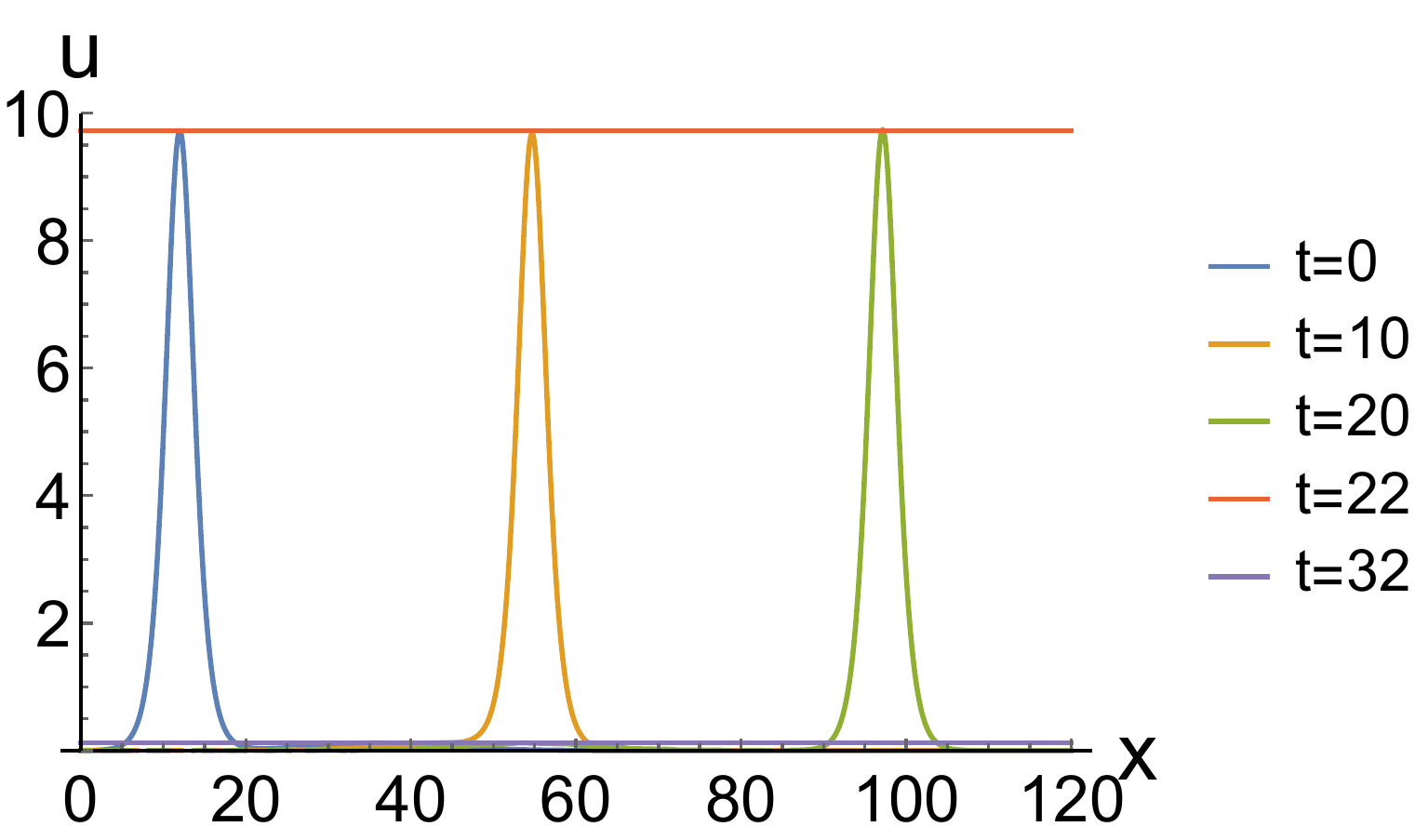}}
    \caption{Interactions between solitons possessed by the BBM equation: (i) perfect merging interaction cases of with (a) $b_1=0.5, b_2=0.214299$ and (b) $b_1=0.75, b_2=0.280551$; (ii) peak merging interaction cases with (c) $b_1=0.55, b_2=0.45$ and (d) $b_1=0.65, b_2=0.35$; (iii) runover merging interaction cases with (e) $b_1=0.8, b_2=0.2$ and (f) $b_1=0.9, b_2=0.1$.}\label{fig2}
\end{figure*}

The most interesting phenomena happen when two input solitary waves merge most. Different
combinations of the input solitary waves will lead to different interaction shapes at the moment of their
maximum merging. These shapes at the maximum merging moment can be categorised into two groups,
including the perfect merging, as shown in Fig.~\ref{fig1} (a-b), and the imperfect merging with two types of interactions: the peak merging and the run-over merging, as shown in Fig.~\ref{fig1} (c-d) and Fig.~\ref{fig1} (e-f), respectively.

If the widths of the two incoming solitons have the relationship of $b_1 = 0.5, b_2 = 0.214299$ or
$b_1 = 0.75, b_2 = 0.280551$, they will merge perfectly (only stretched), as shown in Figs.~\ref{fig1} (a-b) and~\ref{fig2} (a-b). In
Fig.~\ref{fig2} (a-b), we can see that the blue curve represents the incoming solitons coming from the left, the green
wave the merged shape and the purple curve the solitons generated by the interaction. Obviously, the
input wave with higher amplitude propagates much faster than the lower input wave, and likewise for
the output wave.

If the widths of the two solitons are close to each other, such as $b_1 = 0.55, b_2 = 0.45$ or $b_1 = 0.65, b_2 = 0.35$,
the two waves will never merge completely. There can be either the double peak or
the single peak of the merged wave, as shown in Figs.~\ref{fig1} (c-d) and~\ref{fig2} (c-d). As demonstrated in Fig.~\ref{fig2} (c-d), the faster
peak approaches the slower peak to interact, and it reduces height as it transfers energy to the slower
wave until both amplitudes are the same at the maximum merging moment. After that moment, the
original slower wave grows higher, becomes faster, and moves away.

If the widths of the two solitons are quite different, for example, $b_1 = 0.8, b_2 = 0.2$ or $b_1 = 0.9, b_2 = 0.1$,
the higher wave will run straightforwardly over the lower wave as shown in Figs.~\ref{fig1} (e-f) and~\ref{fig2} (e-f). As
presented in Fig.~\ref{fig2} (e-f), the high thin wave sits on top of the short wide wave which creates a shape
reminiscent of a traffic cone at the point of maximum contact.

The results shown in Figs.~\ref{fig1} and \ref{fig2} were designed as the benchmark for our conserved quantity method, as demonstrated in the following parts of this section. The numerical results in Figs.~\ref{fig1} and \ref{fig2} were obtained by using the pseudo-spectral method, and the computational time required for each calculation case when using this numerical method is at least 10 seconds (Lenovo P720).

\subsection{Perfect merging}

\begin{figure*}[t]
    \centering
    \includegraphics[width=0.45\textwidth]{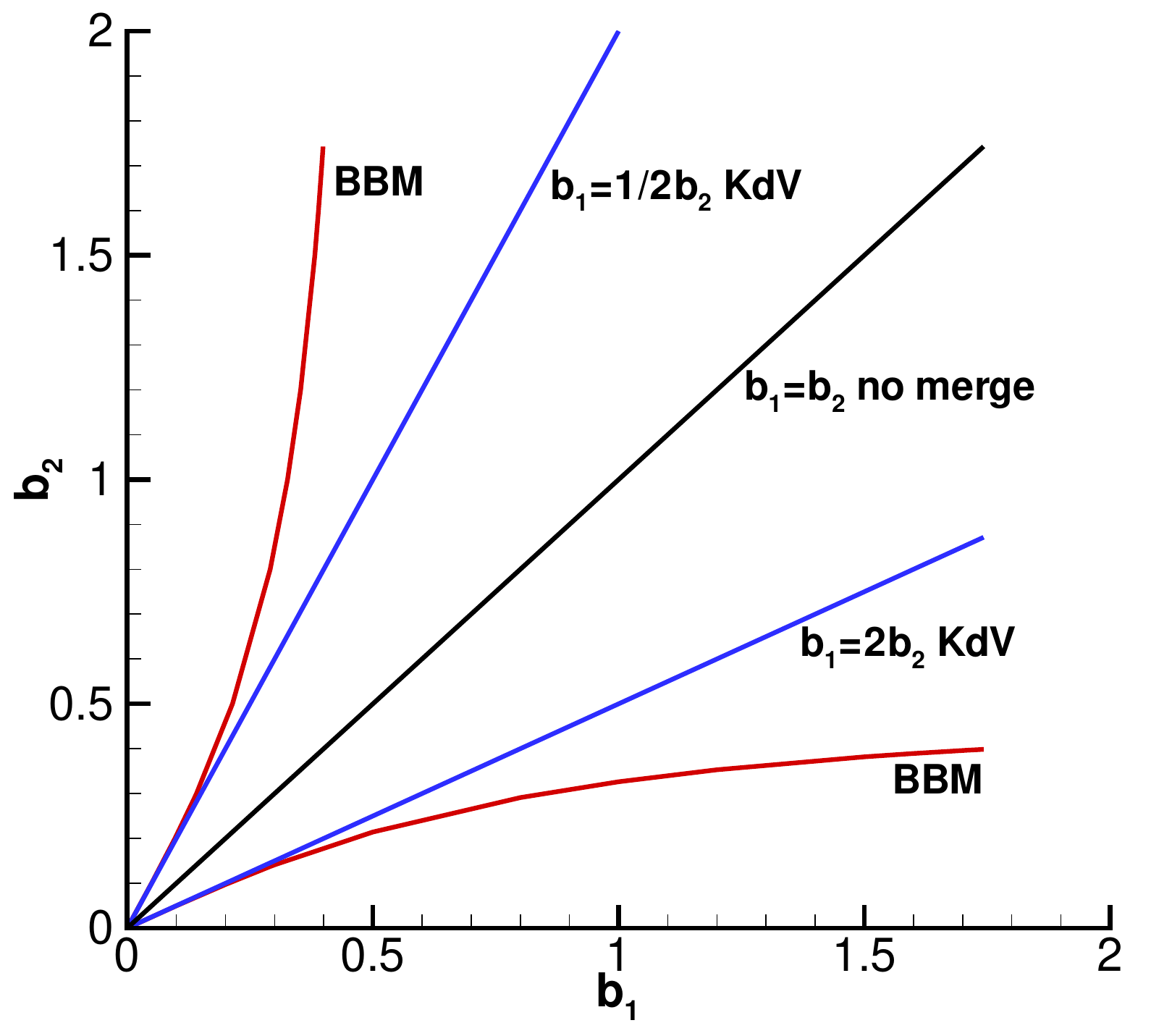}
    \caption{Combination of inverse widths, $b_1$ and $b_2$, of two solitons of the BBM equation to have perfect merging.}\label{fig3}
\end{figure*}

Let us start with a special case, the perfect merging when the interaction wave shape at the
maximum merging moment is the same as the incoming solitary waves except for the width and the
amplitude. Such a case needs a particular ratio of parameter $b$ between the two input waves. As such,
when we know the profile of one incoming wave, we need to find what the other input wave look like
and the shape of the interaction wave spontaneously.

Since the BBM equation is modified from the KdV equation, the shape of the merged wave is
known to be in the form of $\sech^2(x)$. The conserved quantities of two solitary waves can be found
by using the simple integrations listed in Table~\ref{tab1}. The merged wave shape $U$ in the perfect merging case
is assumed to be a stretching soliton defined by the unknown constant $m$, which is multiplied by the
unknown inverse width $B$:
\begin{equation}
U=12m B^2 \sech^2(\frac{Bx}{\sqrt{1+4B^2}}).\label{eq0005}
\end{equation}
Here, the time dependence is dropped as we are interested in the wave profile at the maximum merging
instance.

\begin{table*}[t]
\centering
\caption{\label{tab1} The conserved quantities of  two solitary waves and perfect merging wave shape.}
\begin{tabular}{ c | c | c }
\hline
Conserved quantity    &Solitary wave                                   &Perfect merging                    \T\B                  \\
\hline
$I_1$                 &$24b\sqrt{1+4b^2}$                              &$24Bm\sqrt{1+4B^2}$                  \T                \\
\rule{0pt}{5ex}
$I_2$                 &$\displaystyle{\frac{192b^3(5+24b^2)}{5\sqrt{1+4b^2}}}$        &$\displaystyle{\frac{192B^3m^2(5+24B^2)}{5\sqrt{1+4B^2}}}$          \\
\rule{0pt}{5ex}
$I_3$                 &$\displaystyle{\frac{2304b^5(3+16b^2)}{5\sqrt{1+4b^2}}}$       &$\displaystyle{\frac{2304B^5m^2[4m(1+4B^2)-1]}{5\sqrt{1+4B^2}}}$ \B   \\
\hline
\end{tabular}
\end{table*}

\begin{table*}
\centering
\caption{\label{tab2} The conserved quantities of  two solitary waves and merged wave shape.}
\resizebox{1.0\textwidth}{!}{
\begin{tabular}{c | c | c | c | c | c}
\hline
Input wave $b_1$ (known)    &Input wave $b_2$ (unknown)   & $m$   &B   &$b_1/b_2$    &Max height adjusted error\\
\hline
0             &0                      &0                       &0                     &0              &-        \\[1mm]
0.01          &0.005                  &3.0002                  &0.005                 &2              &-        \\[1mm]
0.05          &0.024956               &3.004980                &0.025006              &2.004          &-        \\[1mm]
0.1           &0.049651               &3.019687                &0.050045              &2.014          &0.02     \\[1mm]
0.2           &0.097254               &3.075278                &0.100266              &2.056          &0.01     \\[1mm]
0.3           &0.141090               &3.158142                &0.150526              &2.126          &0.002    \\[1mm]
0.4           &0.180170               &3.258113                &0.200486              &2.220          &0.004    \\[1mm]
0.5           &0.214299               &3.366288                &0.249857              &2.333          &0.01     \\[1mm]
0.6           &0.243797               &3.476106                &0.298501              &2.461          &0.03     \\[1mm]
0.7           &0.269202               &3.583261                &0.346408              &2.600          &0.02     \\[1mm]
0.8           &0.291088               &3.685184                &0.393642              &2.748          &0.03     \\[1mm]
0.9           &0.309986               &3.780515                &0.440297              &2.903          &0.02     \\[1mm]
1             &0.326357               &3.868682                &0.486471              &3.064          &0.06     \\[1mm]
1.5           &0.381896               &4.207337                &0.712934              &3.928          &-        \\[1mm]
1.7           &0.396004               &4.303260                &0.802466              &4.293          &-        \\[1mm]
1.74          &0.398450               &4.320307                &0.820341              &4.367          &-        \\[1mm]
1.7427        &0.398611               &4.321434                &0.821547              &4.371932       &-        \\[1mm]
1.74270876    &0.398612               &4.321437                &0.821555              &4.371943       &-        \\
\hline
\end{tabular}
}
\end{table*}

As a test by using the conserved quantities to find the maximum wave profile of the perfect merging,
we assume $b_1$ for the first incoming solitary wave is known, and we would like to find the value of
$b_2$ of the second wave to make the perfect merging happen with the first wave, as well as
the maximum wave profile of such an interaction. In this case, we have three unknowns: $b_2$, $m$ and $B$.
Correspondingly, we have three equations that relate the three conservative quantities, $I_1,I_2$ and $I_3$.
From Table \ref{tab1}, by forcing the sum of the input quantities equal to the combined quantities, we have
\begin{align}
&I_1: \quad 24b_1\sqrt{1+4b_1^2}+24b_2\sqrt{1+4b_2^2}=24Bm\sqrt{1+4B^2},\label{eq0006} \\
&I_2: \quad \frac{192b_1^3(5+24b_1^2)}{5\sqrt{1+4b_1^2}}+\frac{192b_2^3(5+24b_2^2)}{5\sqrt{1+4b_2^2}}=\frac{192B^3m^2(5+24B^2)}{5\sqrt{1+4B^2}},\label{eq0007} \\
&I_3: \quad \frac{2304b_1^5(3+16b_1^2)}{5\sqrt{1+4b_1^2}}+\frac{2304b_2^5(3+16b_2^2)}{5\sqrt{1+4b_2^2}}=\frac{2304B^5m^2[4m(1+4B^2)-1]}{5\sqrt{1+4B^2}}.\label{eq0008}
\end{align}
The above three equations can be solved effectively by using one simple function in some symbolic mathematical software packages, such as Mathematica, Maple or Python with SymPy.

\begin{figure*}[!t]
    \centering{}
    \subfloat[]{\includegraphics[width=0.45\textwidth]{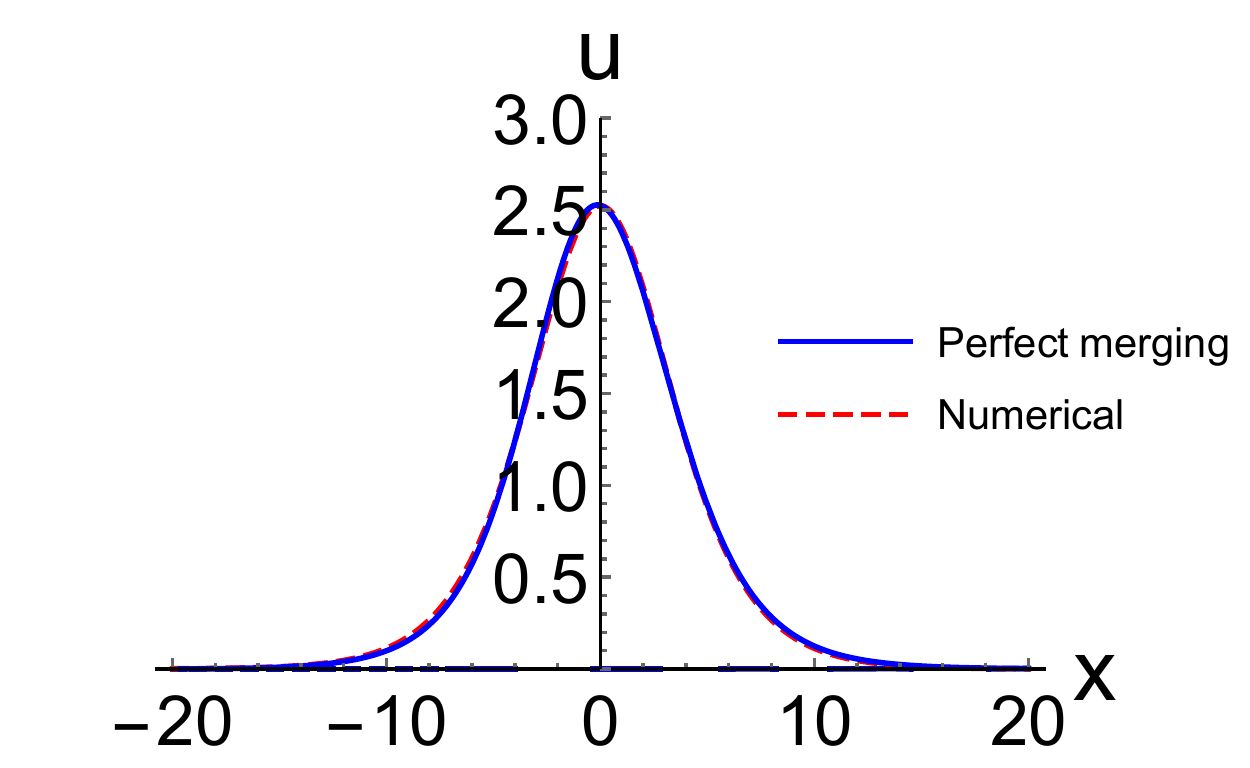}} \quad\quad
    \subfloat[]{\includegraphics[width=0.45\textwidth]{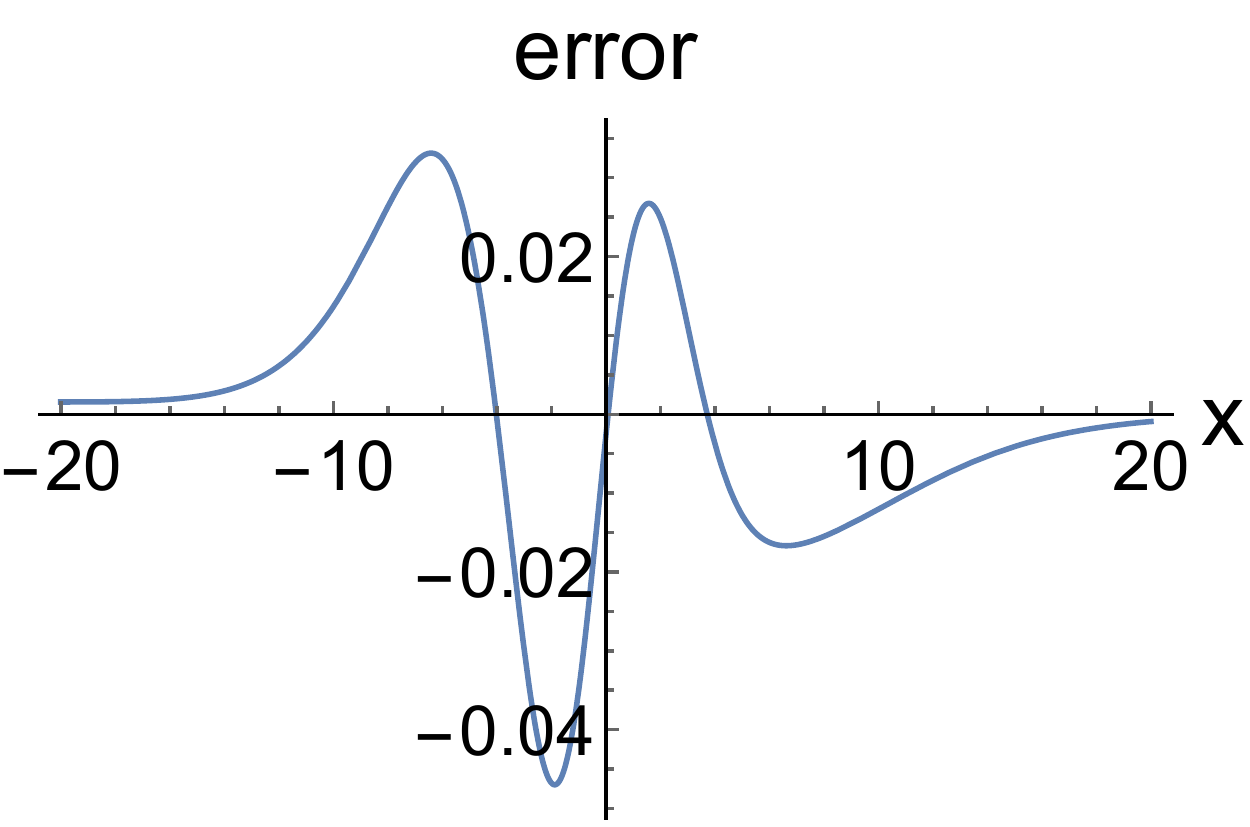}} \\ 
    \subfloat[]{\includegraphics[width=0.45\textwidth]{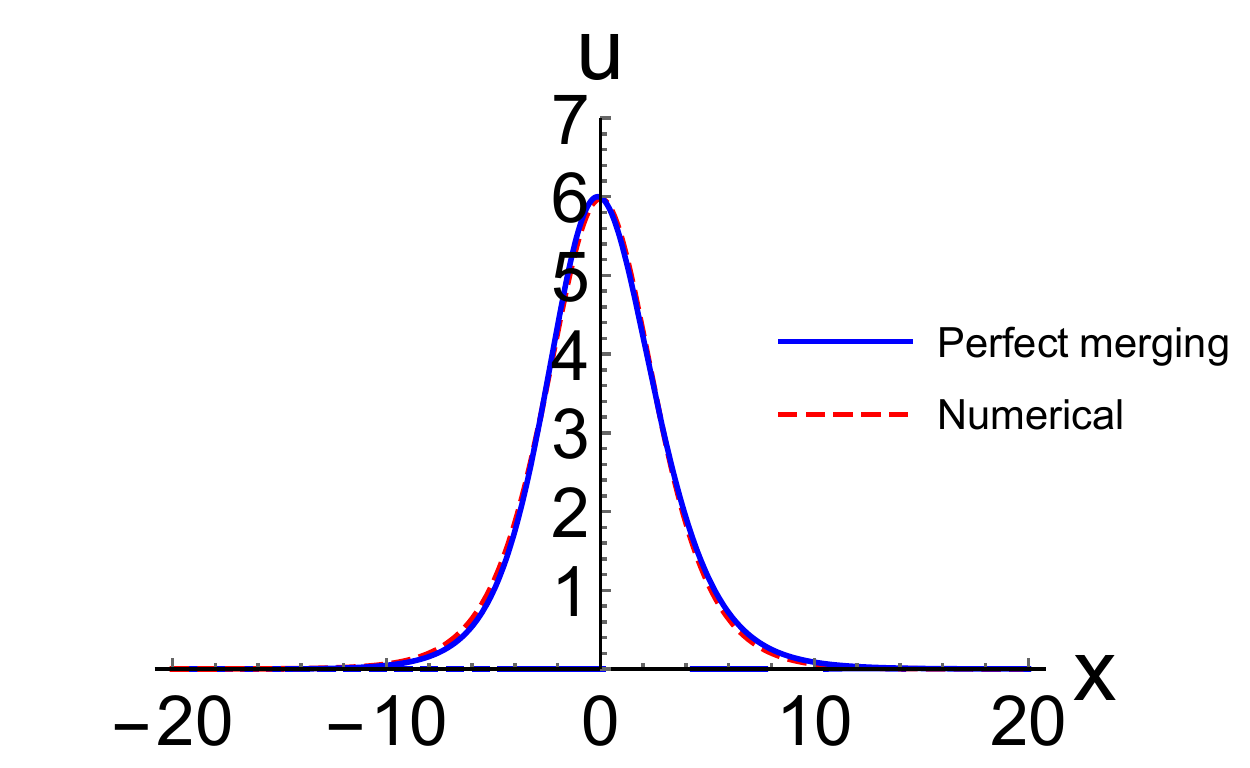}} \quad\quad
    \subfloat[]{\includegraphics[width=0.45\textwidth]{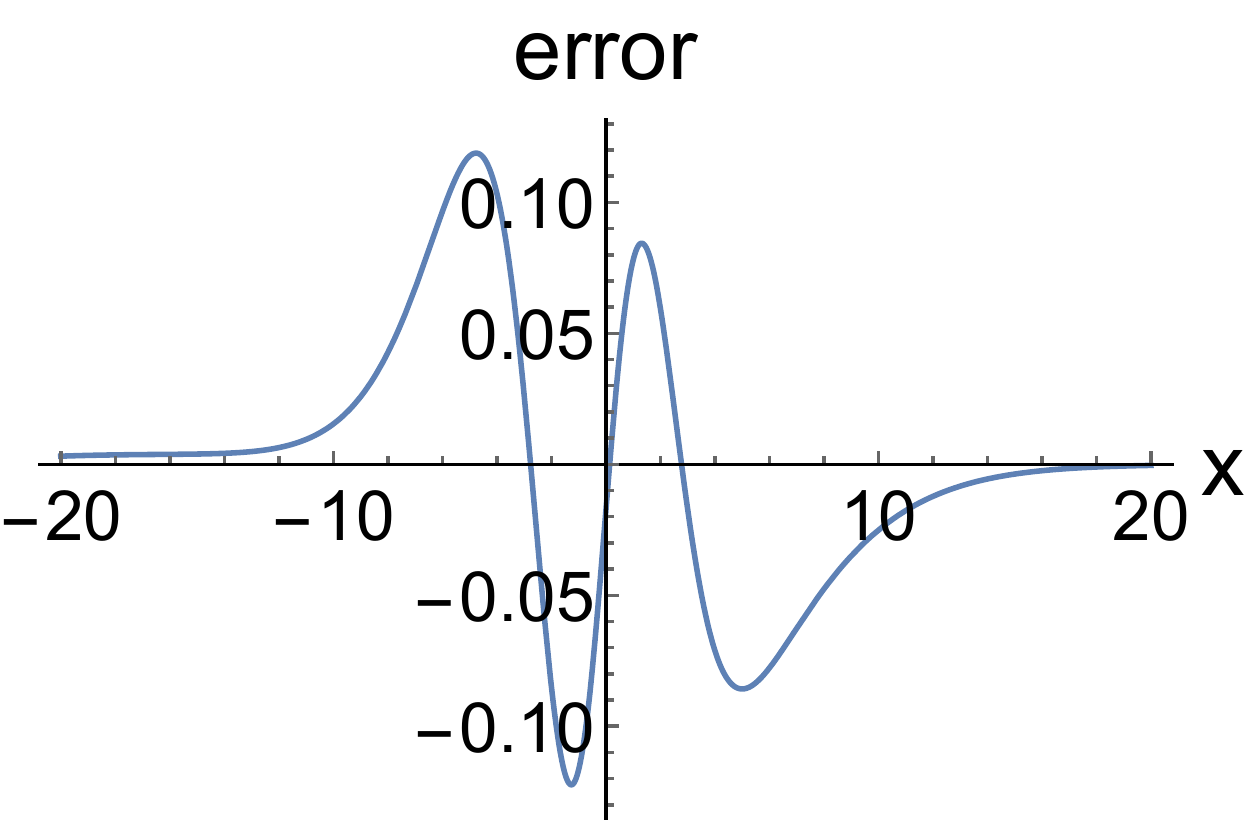}} \\
    \caption{Comparisons (a, c) and (b, d) difference (b, d) of the results between the conserved quantity method and numerical method for the perfect merging of two soliton waves of the BBM equation with (a, b) $b_1=0.5, b_2=0.214299$ and (c, d) $b_1=0.75, b_2=0.280551$.}\label{fig4}
\end{figure*}

The method demonstrated here allows to find one unknown input wave, which results in a perfect
merging wave with the conservation of mass, momentum and energy. This is the important
application of the conserved quantity method to find the relationship between $b_1$ and $b_2$, which is of
great interests to investigate the interactions between long surface waves with finite amplitude travelling
unidirectionally in a nonlinear dispersive system. The combination of the two input solitons required
for a perfect merging possessed by the BBM equation is listed in Table~\ref{tab2}. Fig.~\ref{fig3} shows the comparison
between the perfect merging due to the solitary waves of the KdV equation with $b_1 = 2b_2$ or $b_1 = 1/2b_2$,
and that of the BBM equation according to the data in Table \ref{tab2}.

The accuracy of the results obtained by the conserved quantities is compared with the numerical
results, as shown in Fig.~\ref{fig4}.
The figure shows the interaction wave profile at the perfectly merging
moment by the conserved quantity method and the numerical method for $b_1 = 0.5, b_2 = 0.214299, m = 3.366288$, $B = 0.249857$ and $b_1 = 0.75, b_2 = 0.280551, m = 3.634984$, $B = 0.370103$.
As we can see, good agreement has been found.

\subsection{Imperfect merging}
Having investigated a special case of the perfect merging, we turn our attentions to general cases of
imperfect merging to find the interaction profile when the profiles of the two incoming waves are known.
We can separate such merging cases into two catalogues: (i) The peak merging when the profiles of
two input solitary waves are similar to each other; and (ii) The run-over merging when the two input
solitary waves are very dissimilar to each other.

\subsubsection{Peak merging}
For the cases when two input solitary waves have similar amplitudes and therefore cannot completely
merge, a single solitary wave form is not enough for the shape of the merged wave. Obviously, the
shape with double peaks is needed. As such, we can set the shape of the interaction wave, $U$, at the
maximum merging moment is in the form of
\begin{equation}
U=A \sech^2\left[B\left(x-\frac{L}{2}\right)\right]+A \sech^2\left[B\left(x+\frac{L}{2}\right)\right],\label{eq0009}
\end{equation}
in which $A$ is the height of equal double peaks due to the symmetry, $B$ the inverse width and $L$ the peak
interval. These three unknowns can be solved in the same manner as demonstrated in the previous
section by using the conserved quantities given in Table \ref{tab1} and Table \ref{tab3}.

\begin{table*}[t]
\centering
\caption{\label{tab3} The conserved quantities of peak merging wave shape.}
\begin{tabular}{ c | c }
\hline
Conserved quantity    &Peak merging                  \T\B                  \\
\hline
$I_1$     &$\frac{4A}{B}$                 \T                \\
\rule{0pt}{5ex}
$I_2$     &$\frac{8A^2\left[-1+15e^{2BL}-15e^{4BL}+e^{6BL}+12e^{2BL}\left(1+e^{2BL}\right)log\left(e^{BL}\right)\right]}{3B\left(-1+e^{2BL}\right)^3}$ \\
\rule{0pt}{5ex}
$I_3$     &$\frac{32A^3\left(-1+e^{2BL}\right)\left(1+26e^{2BL}+306e^{4BL}+26e^{6BL}+e^{8BL}\right)}{15B\left(-1+e^{2BL}\right)^5}$\\[1mm]
          &$-\frac{32A^2B^2\left(-1+e^{2BL}\right)\left(1+116e^{2BL}+486e^{4BL}+116e^{6BL}+e^{8BL}\right)}{5B\left(-1+e^{2BL}\right)^5}$\\[1mm]
          &$+\frac{384A^2e^{2BL}\left(1+e^{2BL}\right)\left[-2Ae^{2BL}+B^2\left(1+10e^{2BL}+e^{4BL}\right)\right]\texttt{log}\left(e^{BL}\right)}{B\left(-1+e^{2BL}\right)^5}$ \B   \\
\hline
\end{tabular}
\end{table*}

\begin{figure*}[!t]
    \centering{}
    \subfloat[]{\includegraphics[width=0.45\textwidth]{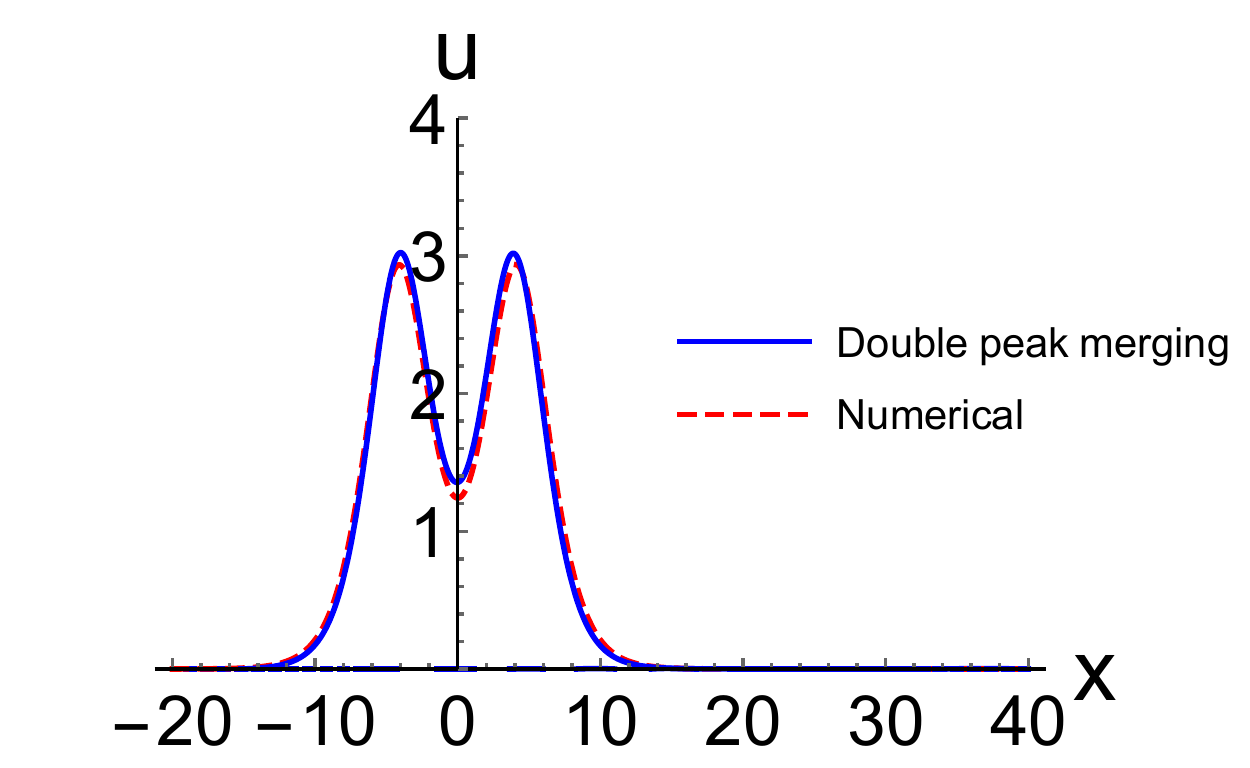}} \quad\quad
    \subfloat[]{\includegraphics[width=0.45\textwidth]{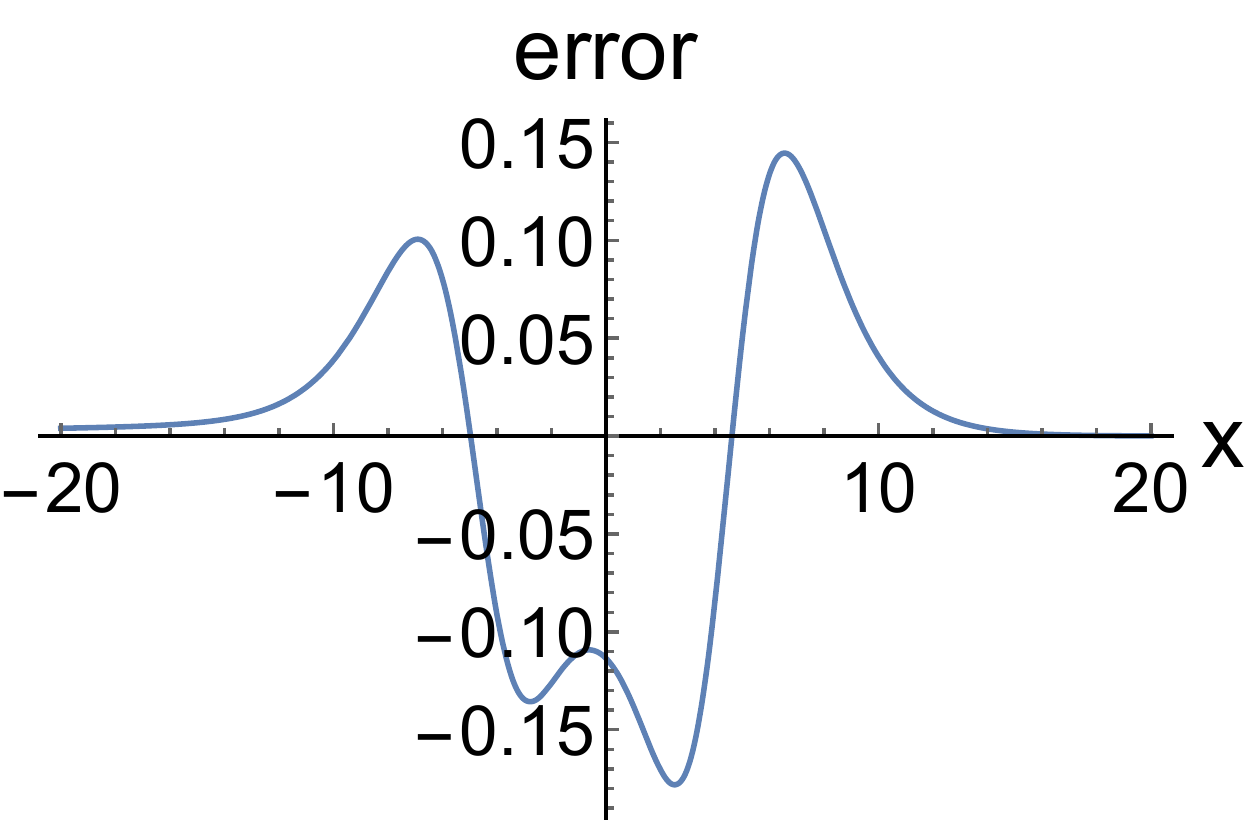}} \\ 
    \subfloat[]{\includegraphics[width=0.45\textwidth]{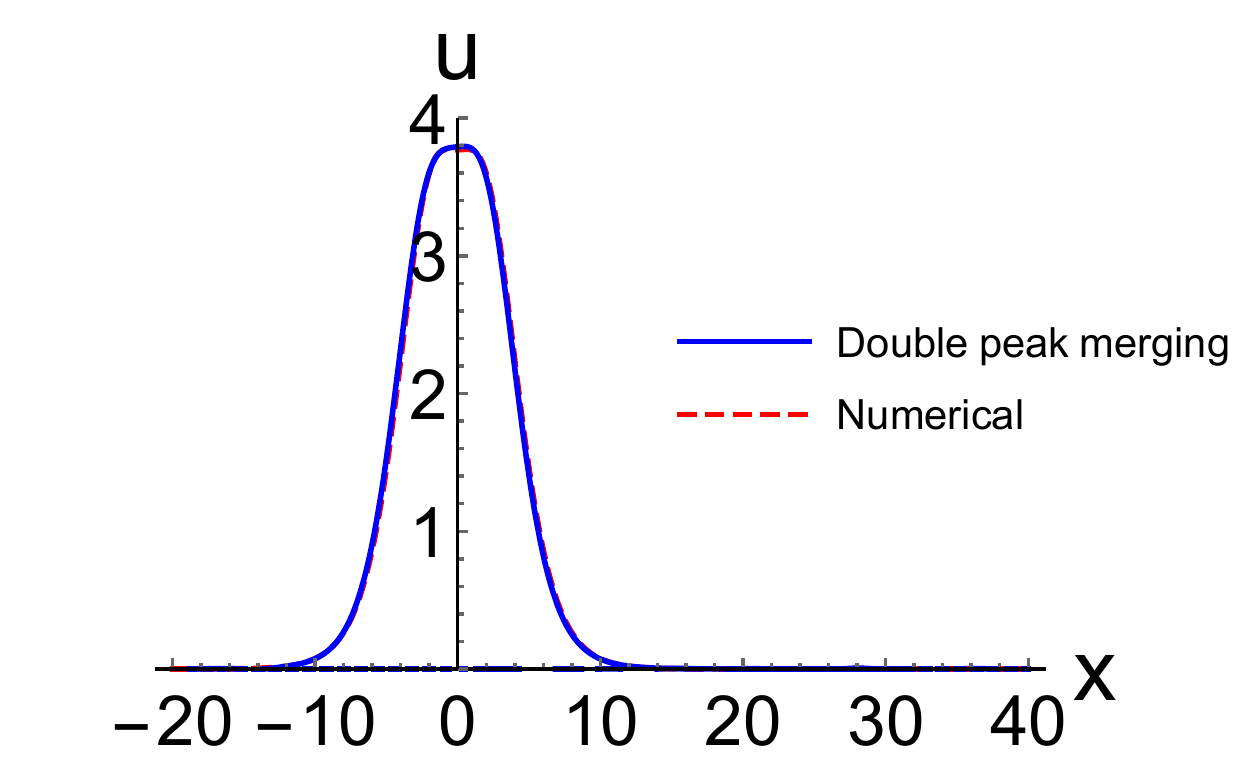}} \quad\quad
    \subfloat[]{\includegraphics[width=0.45\textwidth]{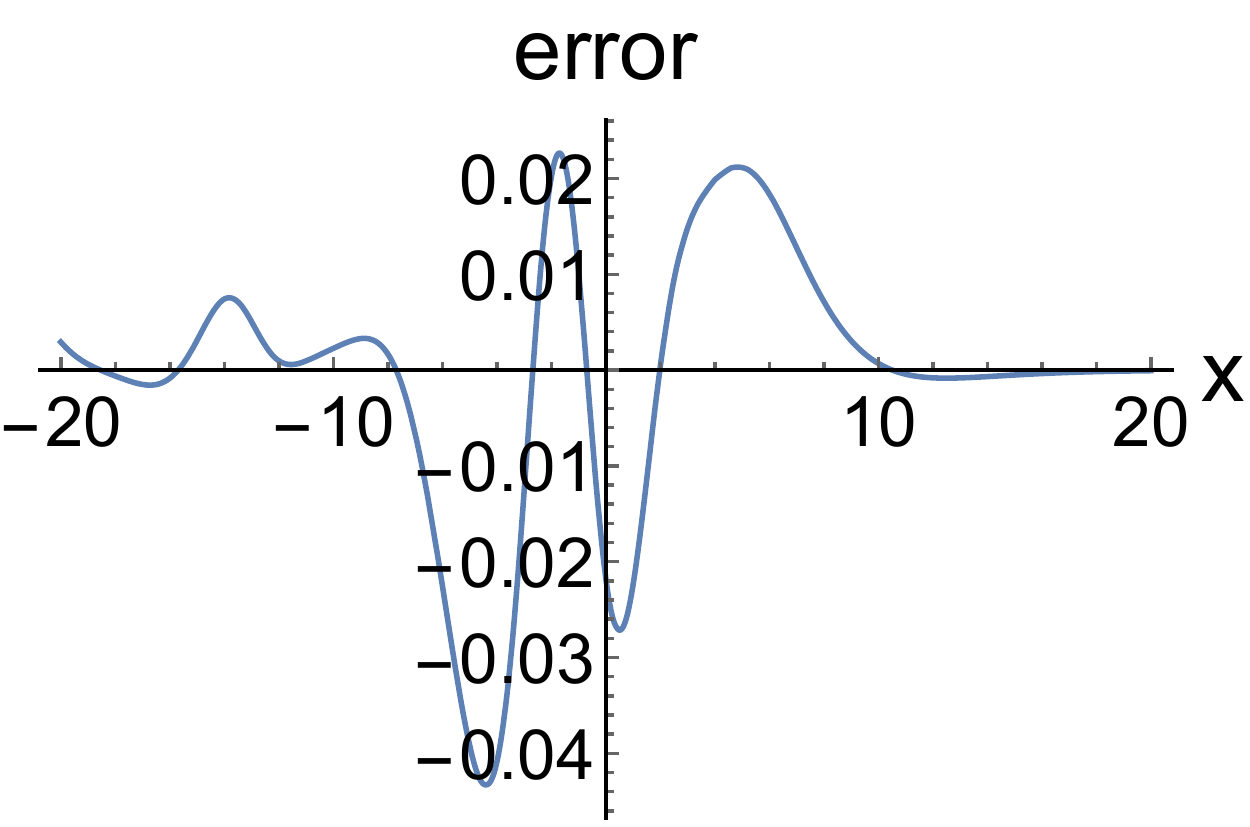}} \\
    \caption{Comparisons (a, c) and (b, d) difference (b, d) of the results between the conserved quantity method and numerical method for the peak merging of two soliton waves of the BBM equation with (a, b) $b_1=0.55, b_2=0.45$ and (c, d) $b_1=0.65, b_2=0.35$.}\label{fig5}
\end{figure*}

To validate our method, we chose the following pairs of $b_1$ and $b_2$. When $b_1 = 0.55$ and $b_2 = 0.45$,
$A = 2.8918826$, $B = 0.338696$ and $L = 8.290023$ are obtained by using the conserved quantity method.
When $b_1 = 0.65$ and $b_2 = 0.35$, $A = 2.883558$, $B = 0.3218314$ and $ L = 4.197039$ are obtained. Compared with the
numerical results, we can see that the shape of the merged wave is almost identical to that by using
the numerical method and the difference are insignificant, as shown in Fig.~\ref{fig5}.
This indicates that good agreement has been found between the results obtained by our conserved quantity
method and the numerical method.

\subsubsection{Run-over merging}
In the case of two solitary waves of BBM equation with very different amplitudes, the fast incoming
wave will run directly over the slow wave, instantly forming a merged shape with a thin peak on top
of a wide base. The shape of the run-over merging wave, $U$, can be written in the form of
\begin{equation}
U=A \sech^{2n}(Bx),\label{eq0010}
\end{equation}
in which $A$ is the height of the merging wave, $B$ the inverse width and $n$ the shape parameter. These three unknowns are solved in the same way by using the conserved quantities given in
Table \ref{tab1} and Table \ref{tab4}.

\begin{figure*}[!t]
    \centering{}
    \subfloat[]{\includegraphics[width=0.45\textwidth]{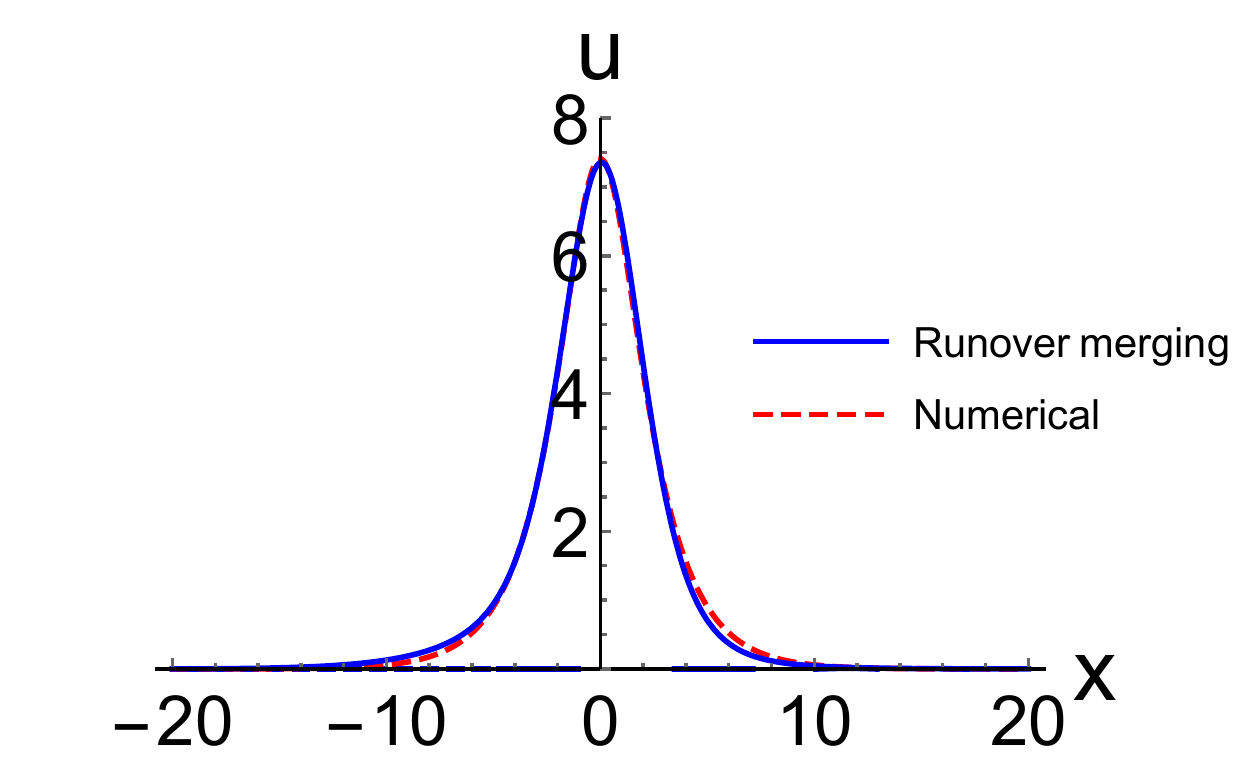}} \quad\quad
    \subfloat[]{\includegraphics[width=0.45\textwidth]{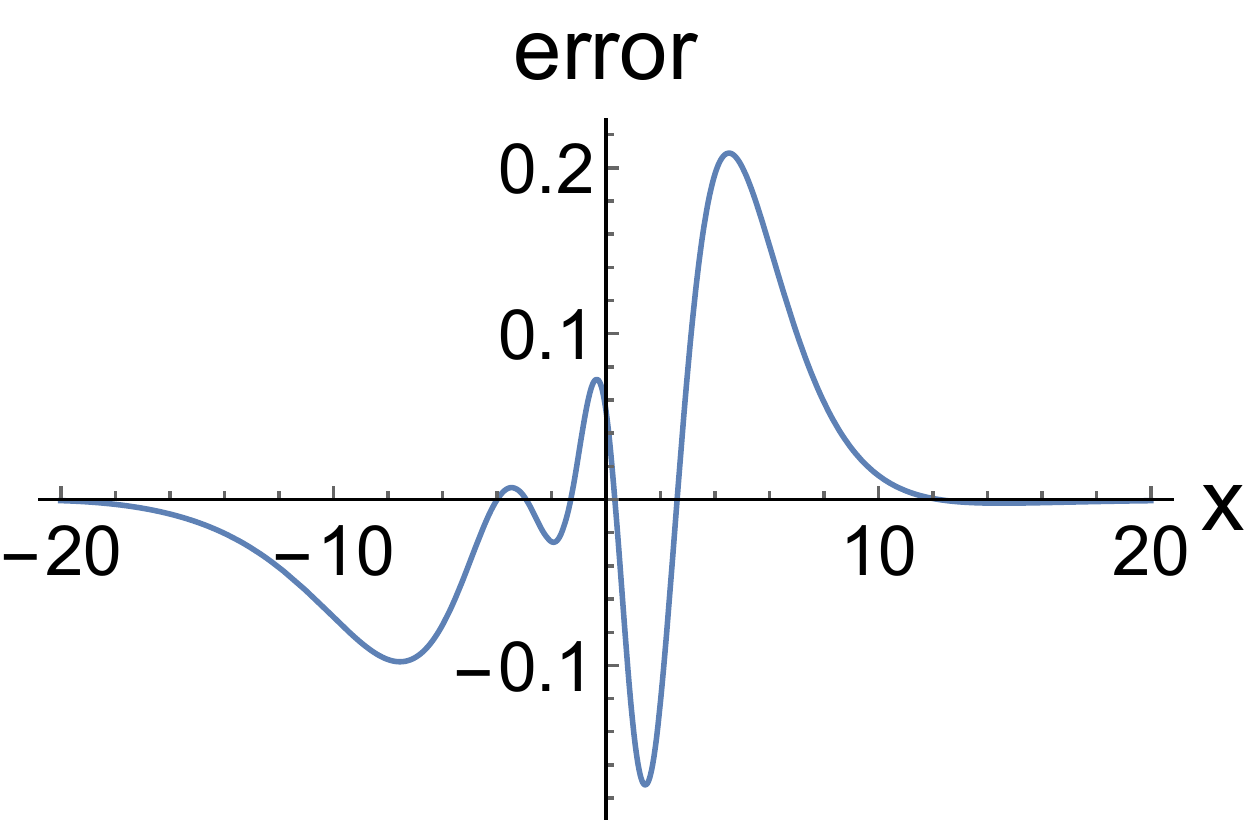}} \\ 
    \subfloat[]{\includegraphics[width=0.45\textwidth]{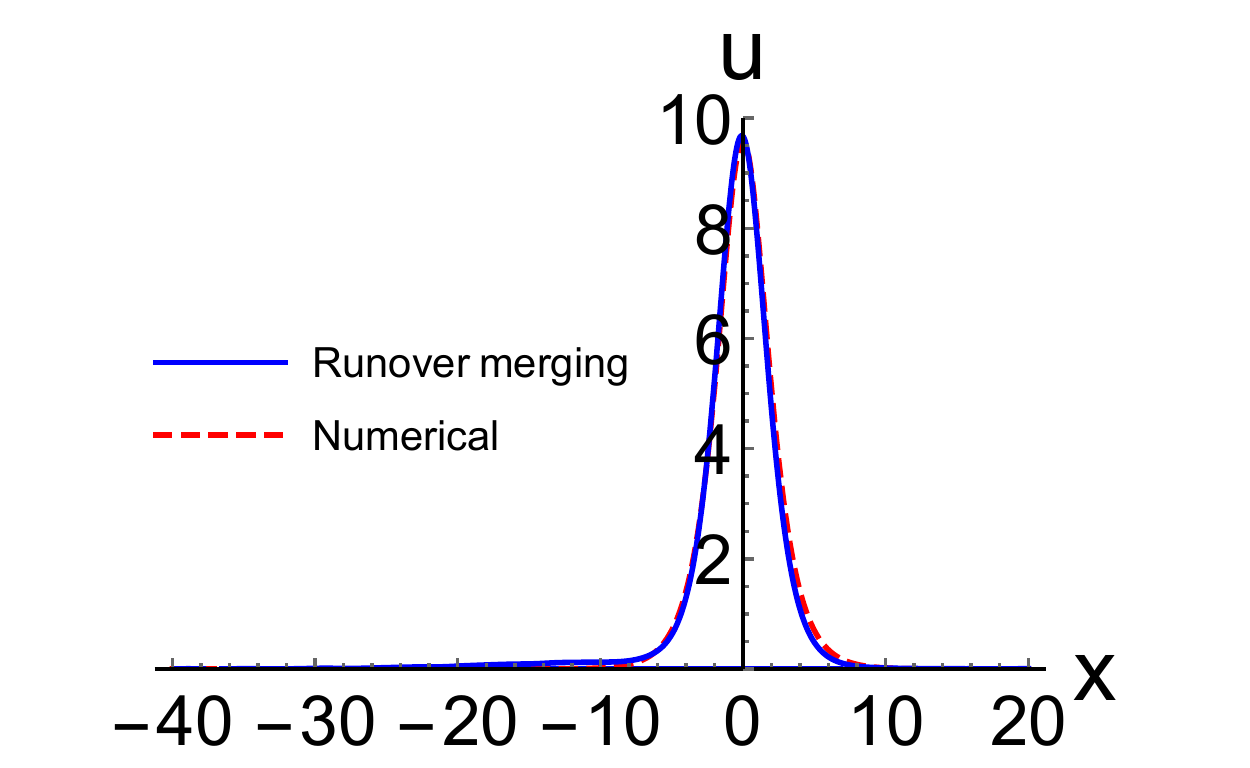}} \quad\quad
    \subfloat[]{\includegraphics[width=0.45\textwidth]{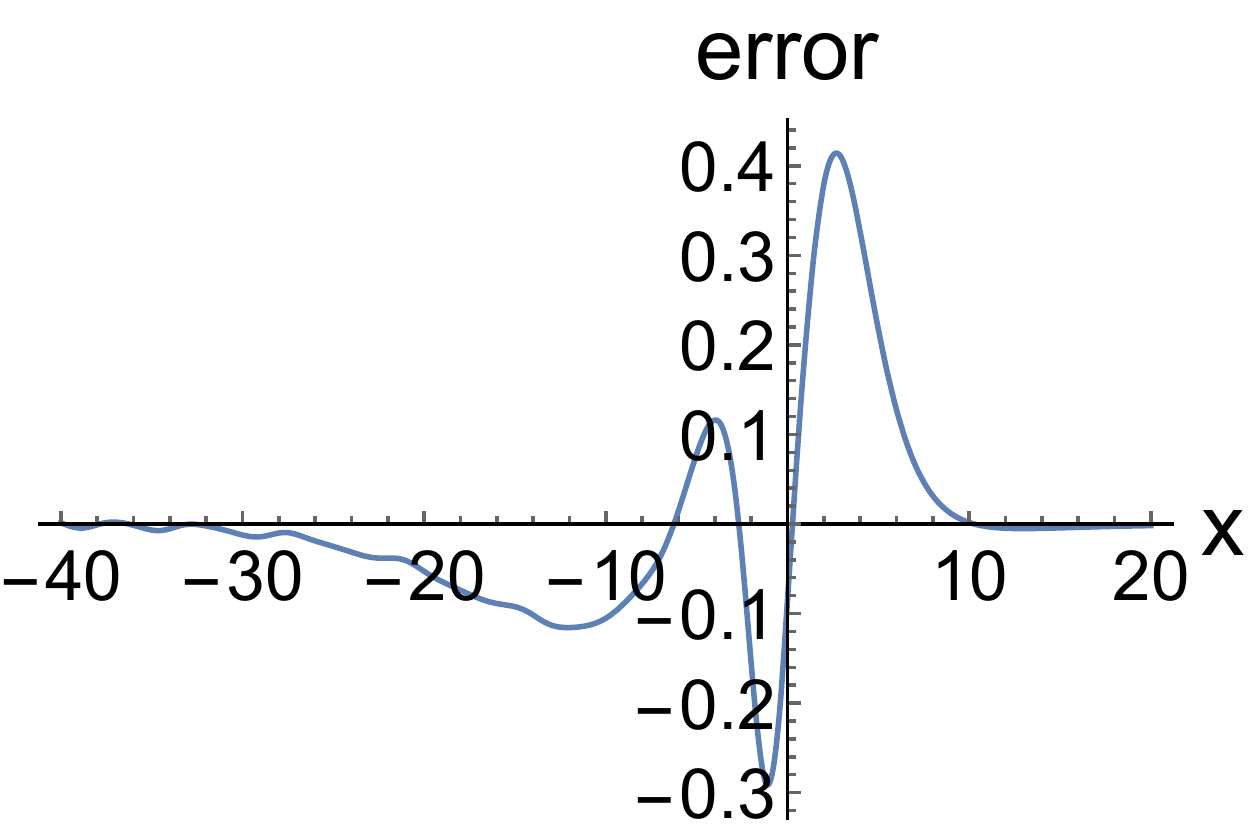}} \\
    \caption{Comparisons (a, c) and (b, d) difference (b, d) of the results between the conserved quantity method and numerical method for the run-over merging of two soliton waves of the BBM equation with (a, b) $b_1=0.8, b_2=0.2$ and (c, d) $b_1=0.9, b_2=0.1$.}\label{fig6}
\end{figure*}

To examine the accuracy of run-over merging results given by the conserved quantity method,
we compare the shape of the interaction wave at the maximum merging moment with that by the
numerical methods. When $b_1 = 0.8, b_2 = 0.2$ and $b_1 = 0.9, b_2 = 0.1$, we found that $A = 7.406503, B = 0.605108, n = 0.45$ and $A = 9.582191, B = 0.516520, n = 0.69$, respectively.
For such pairs of $b_1$ and $b_2$, the shapes of the merged wave obtained by the conserved quantity method are almost identical to those by the numerical method, as shown in Fig.~\ref{fig6}, which validates the accuracy and robustness of the conserved quantity method.

\begin{table*}[t]
\centering
\caption{\label{tab4} The conserved quantities of peak merging wave shape.}
\begin{tabular}{ c | c }
\hline
Conserved quantities    &Runover merging                  \T\B                  \\
\hline
$I_1$     &$\frac{4^n A~_{2}F_{1}(n,2n;1+n;-1)}{Bn}$                \T                \\
\rule{0pt}{5ex}
$I_2$     &$\frac{A^2 \sqrt{\pi} \Gamma(2n)}{B \Gamma\left(\frac{1}{2}+2n\right)}$ \\
\rule{0pt}{5ex}
$I_3$     &$\frac{16^n A^2 (-18Bn) ~_{2}F_{1}(2n,2+4n;1+2n;-1)}{3}$\\[1mm]
          &$+\frac{16^n A^2}{3} \frac{A4^n  ~_{2}F_{1}(3n,2+6n;1+3n;-1)}{B n}$\\[1mm]
          &$-\frac{16^n A^2}{3}\frac{18B n^2 ~_{2}F_{1}(2(1+n),2+4n;3+2n;-1)}{1+n}$\\[1mm]
          &$+\frac{16^n A^2}{3} \frac{72Bn^2 ~_{2}F_{1}(1+2n,2+4n;2(1+n);-1)}{1+2n}$\\[1mm]
          &$\frac{16^n A^2}{3} \frac{2A 34^n ~_{2}F_{1}(1+3n,2+6n;2+3n;-1)}{B+3Bn}$\\[1mm]
          &$\frac{16^n A^2}{3} \frac{A 34^n ~_{2}F_{1}(2+3n,2+6n;3(1+n);-1)}{2B+3Bn}$ \B   \\
\hline
\end{tabular}
\end{table*}

\section{Conclusions}

In this paper, we introduce a simple, robust and efficient method, the conserved quantity method,
to study the maximum interaction process of two solitary waves possessed by the BBM equation. By
using this method, the main advantage is that there is no need to solve the nonlinear partial differential equation (BBM equation)
explicitly, and such a conserved quantity method can successfully give an accurate approximation to the
size and shape of the merged wave. To demonstrate our method, we showed three types of interaction phenomena of two incoming
solitary waves of the BBM equation at the maximum merging moment, including the perfect merging,
the peak merging and the run-over merging, and good agreement is found between the results of our conserved quantity analytical method and the numerical results. Though it is assumed that the dispersion tail is not significant
when using this method, it is a reasonable assumption for the purpose of practical applications, such as
offshore and coastal engineering. The method presented in this work can be improved by
adding a parameter to explain the dispersion tail created during the interaction with the cost of solving
a more complex set of equations. Applications of this conserved quantity method can be considered to analyse the interacting phenomena between solitary waves in the areas of shallow water waves, condensed matter physics and nonlinear optics.

\section*{Acknowledgments}
This work is partially funded by the National Natural Science Foundation of China (Grant No. 12002390). Q.S. was supported by the Australian Research Council (ARC) through Grants DE150100169, FT160100357 and CE140100003.

\vspace{1cm}
\begin{tabular}{cp{0.4\textwidth}}
{\bf Nomenclature}& \\
$u$                     & non-dimensional surface elevation                                                \\
$t$                     & non-dimensional time                                                             \\
$x$                     & non-dimensional displacement                                                     \\
$\bar{a}$               & non-dimensional amplitude                                                        \\
$\bar{b}$               & non-dimensional inverse width                                                    \\
$\bar{c}$               & non-dimensional wave celerity                                                    \\
$b$                     & non-dimensional inverse width                                                    \\
$I_1$                   & conserved quantity of mass                                                       \\
$I_2$                   & conserved quantity of momentum                                                   \\
$I_3$                   & conserved quantity of energy                                                     \\
$b_1$                   & inverse width of input wave one                                                  \\
$b_2$                   & inverse width of input wave two                                                  \\
$U$                     & merged wave shape                                                                \\
$m$                     & constant of merged wave                                                          \\
$n$                     & index                                                                            \\
$A$                     & amplitude of merged wave                                                         \\
$B$                     & inverse width of merged wave                                                     \\
$L$                     & peak interval of merged wave                                                     \\
\end{tabular}
\vspace{1cm}

 \bibliographystyle{elsarticle-num} 
 \bibliography{mybibfile}





\end{document}